\documentclass[preprint,aps,showkeys,showpacs]{revtex4}
\usepackage[english]{babel}
\usepackage{graphicx}
\usepackage{graphics}
\usepackage{amsmath}
\usepackage{dcolumn}
\usepackage{amssymb}
\usepackage{bm}

\begin{document}

\title{Quantization and stability of bumblebee electrodynamics}

\author{C. A. Hernaski}
\email{chernask@indiana.edu}
\affiliation{Indiana University Center for Spacetime Symmetries,
Bloomington, Indiana 47405, USA}

\begin{abstract}
The quantization of a vector model presenting spontaneous breaking of Lorentz symmetry in flat Minkowski spacetime is discussed. The Stueckelberg trick of introducing an auxiliary field along with a local symmetry in the initial Lagrangian is used to convert the second-class constraints present in the initial Lagrangian to first-class ones. An additional deformation is employed in the resulting Lagrangian to handle properly the first-class constraints, and the equivalence with the initial model is demonstrated using the BRST invariance of the deformed Lagrangian. The framework for performing perturbation theory is constructed, and the structure of the Fock space is discussed. Despite the presence of ghost and tachyon modes in the spectrum of the free theory, it is shown that one can implement consistent conditions to define a unitary and stable reduced Fock space. Within the restricted Fock space, the free model turns out to be equivalent to the Maxwell electrodynamics in the temporal gauge.
\end{abstract}

\maketitle

\section{Introduction}

A great amount of effort has been employed in the construction of
a consistent quantum theory of gravity. This relies mainly on theoretical
grounds, since direct experimental evidence from the Planck scale $\simeq10^{19}$ GeV, where such a theory plays a major
role, is presently unattainable. However, it may happen that
these fundamental theories can provide some key signals that
our currently low-energy-scale experiments are able to detect. Since CPT and Lorentz symmetry are among the tenets of our present
understanding of nature at the fundamental level, minor deviations
from these symmetries could be detected in low-energy experiments. In this way,
CPT and Lorentz violation is one of these key signals and, interestingly,
it may occur in many of the candidates for fundamental theories, like
string theory \cite{string1, string2}, loop quantum gravity \cite{lqg}, noncommutative field theories \cite{non comm}, and nontrivial spacetime topology \cite{non trivial top}.

The effective field theory that accounts for the possible deviations
of the known physical phenomena due to 
Lorentz and CPT violations is the Standard-Model Extension (SME) \cite{coll-kost, alan-gravity}. In this framework,
the usual Lagrangians of the Standard Model (SM) of the elementary
particles of physics and of Einstein's general theory of relativity
are supplemented by Lorentz-violating (LV) operators. In the nongravitational
sector of the SME, these LV operators are constructed by considering all SM operators contracted with LV tensorial
coefficients in a coordinate-invariant way. The gravity sector, in
turn, follows the same idea, but considering diffeomorphism tensors
instead of the SM operators.

The LV coefficients can be
generated in many different ways. One particularly elegant and generic one
is through spontaneous Lorentz and CPT violation \cite{{string1, SSB lorentz}}.
In this case, along with Lorentz and CPT violation,
other important consequences can arise, like the appearance of
Nambu-Goldstone (NG) and Higgs modes. Unlike the effective framework
provided by the SME, the properties of these modes are, in general,
model dependent and cannot be completely discussed without knowledge
of the underlying fundamental theory. However,
in many cases, some features of the propagation of these modes can
be discovered in a model-independent way. In Ref. \cite{gravity}, for instance, the effects of the NG modes on the metric
field are taken into account using the coordinate invariance requirement.
In the work of Ref. \cite{bluhm}, some general conclusions about the
fate of the NG modes are also obtained without considering any particular
theory. It is also worth mentioning that, unlike the nongravitational
SME sector, the gravity sector needs to take into account the NG modes
to keep its consistency as is shown in Ref. \cite{alan-gravity}.

Being an effective model, the SME is expected to apply at low energies up
to some characteristic energy scale frequently related to the Planck
scale, and for this reason, it is unsurprising that some inconsistencies
can arise in the analysis of some phenomena if the typical energy
scale under consideration is pushed to arbitrarily large values.
Concerning the photon sector of the SME, the works in Ref. \cite{schreck manoel} investigated the subtle issues of microcausality and unitarity. In Ref. \cite{alan-lehnert}, focusing in the fermion sector, the authors conclude that some problems can arise for energies of the order of the Planck scale.  It is also suggested that
contributions coming from the extra modes, due to the spontaneous
symmetry-breaking mechanism in the fundamental theory, could help
in the consistency of the models. To
better understand the role of the NG modes in the problem of the stability
and causality it seems relevant to consider the quantization of models
presenting spontaneous Lorentz and CPT violation and
try to extract from them some general features.

The most studied LV models contemplating the role of the extra modes
arising from the spontaneous symmetry-breaking mechanism involve
the vacuum condensation of a vector field. These are called ``bumblebee
models'' and were discussed in curved and flat spacetimes \cite{string1, gravity, bluhm, stability bumblebee, bumblebee}. Besides
the NG excitations, other modes can also appear, like massive ones
called Higgs modes and Lagrange multiplier modes. The propagation
properties of all the dynamical modes depend on the form of the kinetic
and potential term considered in the Lagrangian. In Ref. \cite{bluhm}, it
was shown that even when the extra modes do not propagate, they can
give interesting and measurable contributions to the Coulomb and Newton
potentials.

Bumblebee models have been extensively investigated not only as
toy models to probe the role of the several excitations originated from the
Lorentz symmetry-breaking mechanism, but also as an alternative to
the $U\left(1\right)$ gauge theory in the consistent description
of the photon. In this case, the masslessness of the photon is unrelated
to the invariance of the system under a local symmetry, but is instead related to its
identification as a NG mode. Surprisingly, with some assumptions,
these LV vector models turn out to be equivalent to the Maxwell electromagnetism
in a special nonlinear gauge. Actually, a very interesting model
considered by Nambu \cite{nambu} already described the photon as a NG mode
due to spontaneous Lorentz violation. In Nambu's
model, Lorentz violation is introduced by choosing a nonlinear
gauge condition and inserting it directly into the Maxwell Lagrangian coupled
to a conserved current. However, unlike the bumblebee models,
Lorentz violation in Nambu's model is unphysical, since it is a
consequence of a special gauge choice. 

This work establishes a suitable formalism for the canonical
quantization of a particular bumblebee model with a Maxwell-type kinetic
term and a smooth quartic potential responsible for triggering the spontaneous
breaking of Lorentz symmetry. This model was introduced by Kosteleck\'{y}
and Samuel (KS) in Ref. \cite{string1} and was investigated in Refs. \cite{bluhm, stability bumblebee}.
Besides the massless NG modes, it propagates a massive tachyonic excitation
leading to instabilities. However, it will be shown that one can consistently choose a region of the phase space of the solutions where the tachyon does not propagate, and within this phase space slice the model is classically equivalent to the Maxwell theory in a nonlinear gauge. Since
the model has second-class constraints, the most direct way to apply
the methods of canonical quantization is through Dirac's method
of the quantization of constrained systems. However, to avoid the difficulties of Dirac's method,
this work makes use of the Stueckelberg method. It consists of
the enlargement of the field content along with the introduction of
a local symmetry in the Lagrangian to turn the second-class constraints into first-class
ones. The first-class constraints, in turn, are handled with the usual
procedure of quantization of gauge theories.
First, a gauge-fixing term will be introduced in the gauge-invariant Lagrangian, and it will be shown that the new regular Lagrangian is BRST invariant, resulting in its equivalence to the KS model. Truncating the Lagrangian up to quadratic terms, the basic components to perform a systematic quantum analysis of the model are constructed. These include the derivation of the dispersion relations of the propagating modes, the subtle Fourier-mode expansion of the free fields, and the correct identification of the creation and annihilation operators as well as their algebra. The perturbative conditions for the absence of negative-norm states and tachyonic excitations are also discussed.  The resulting free model is tachyon free with a positive-normed Fock space that coincides with that of the Maxwell electrodynamics in the temporal gauge. To test the consistency of the treatment, the analysis of the stability of the free model is made and compared with the classical discussion of Ref. \cite{stability bumblebee}.

This paper is organized as follows: Section \ref{sec:Spontaneous-Lorentz-symetry}
reviews the main classical properties of the KS model.
The implementation of the perturbation
analysis and application of the Stueckelberg method to the KS model is discussed in Sec. \ref{sec:Perturbation-analysis-and}. In Sec. \ref{sec:Fourier-modes-expansion},
the Fourier-mode expansion of the fields and the construction
of the extended Fock space of the deformed KS model are performed. The conditions
for the absence of negative-norm states and the stability of the free model are discussed
in Sec. \ref{sec:Stability}. Finally, in Sec. \ref{sec:Summary},
the results are summarized. In Appendix \ref{appendix a}, the BRST invariance of the full proposed Lagrangian is demonstrated. Appendix \ref{appendix b} presents some technical calculations concerning the
Fourier expansion of the fields.

\section{Spontaneous Lorentz symmetry violation and classical stability of
the Kosteleck\'{y}-Samuel model\label{sec:Spontaneous-Lorentz-symetry}}

The starting point is the specific KS model with a smooth quartic potential:
\begin{equation}
\mathcal{L}_{KS}=-\frac{1}{4}B_{\mu\nu}B^{\mu\nu}-\frac{\kappa}{4}\left(B_{\mu}B^{\mu}-b^{2}\right)^{2}-B_{\mu}J^{\mu},\label{eq:Lagrangian}
\end{equation}
where
\begin{equation}
B_{\mu\nu}=\partial_{\mu}B_{\nu}-\partial_{\nu}B_{\mu}.\label{eq:field strength}
\end{equation}
$\kappa$ is a dimensionless positive constant, and $J^{\mu}$ is an assumed
conserved current composed of matter fields, and it is also the source for the $B_{\mu}$
field. In the present analysis, the dynamics for the matter fields
that compose the current $J^{\mu}$ will be disregarded. $b^{2}$ is a positive constant with dimension
of (mass)$^{2}$, and it will be convenient, for the coming discussions,
to consider it as the quadratic scalar $b^{2}=b_{\mu}b^{\mu}$, formed out
of the constant timelike vector $b_{\mu}$. The first term in the Lagrangian 
(\ref{eq:Lagrangian}) is the usual Maxwell term, so it is invariant
under $U\left(1\right)$ gauge transformations. However, the potential
term, $V=\frac{\kappa}{2}\left(B_{\mu}B^{\mu}-b^{2}\right)^{2}$, breaks
this gauge invariance. One can also see that the minimum of the potential
occurs for $B^{2}=b^{2}$. Therefore, the field $B_{\mu}$ acquires a nonvanishing vacuum
expectation value. This indicates the occurrence of spontaneous
breaking of Lorentz symmetry. The vacuum is degenerate, and any
choice between the possible vacuum states leads to equivalent physical
scenarios. For definiteness, the vacuum state is chosen to be such
that $\left\langle B_{\mu}\right\rangle =b_{\mu}$, where $\left\langle \cdot\right\rangle $
means vacuum expectation value. According to Ref. \cite{fung},
one can classify the potential propagating modes in a spontaneously
symmetry-broken model in five types. For the present purposes, due to the
characteristics of the KS model, only two of them are relevant: the NG
modes, which are massless excitations satisfying the condition $V^{\prime}\left(X\right)=0$,
where the prime means derivative with respect to $X=B_{\mu}B^{\mu}-b^{2}$;
and massive modes that satisfy $V^{\prime}\neq0$. Since $V^{\prime}=\kappa\left(B_{\mu}B^{\mu}-b^{2}\right)$,
the massive mode will be present whenever $\left(B_{\mu}B^{\mu}-b^{2}\right)\neq0$.

As discussed in Ref. \cite{stability bumblebee}, the Hamiltonian associated
with Lagrangian (\ref{eq:Lagrangian}) is unbounded from
below for general field configurations. Nonetheless, initial conditions
can be chosen such that, for such field configurations, the Hamiltonian
remains positive.

The conjugate momenta associated with the $B_{\ensuremath{\mu}}$ fields are defined by
\begin{equation}
\Pi^{\mu}\equiv\frac{\delta\mathcal{L}}{\delta\left(\partial_{0}B_{\mu}\right)}.\label{eq:conjugate momenta def}
\end{equation}
From this definition and from Eqs. (\ref{eq:Lagrangian}) and (\ref{eq:field strength}),
one has 
\begin{equation}
\Pi^{\mu}=-B^{0\mu}.\label{eq:conjugate momenta}
\end{equation}
This immediately shows that only three out of the four components
of $B_{\mu}$ actually propagate. In fact, $\Pi^{0}=0$
is identified as the primary constraint on the phase space of the model. Following
the usual Lagrangian approach, the equations of motion can be derived. They are given
by
\[
\partial_{\mu}B^{\mu\nu}-\kappa\left(B_{\mu}B^{\mu}-b^{2}\right)B^{\nu}=J^{\nu}.
\]
Considering $\nu=0$, one gets the consistency condition for the primary
constraint. The two constraints
\begin{eqnarray}
\phi & = & \Pi^{0}\thickapprox0,\label{eq:primary constraint}\\
\chi & = & \partial_{i}\Pi^{i}-\kappa\left(B_{\mu}B^{\mu}-b^{2}\right)B^{0}-J^{0}\thickapprox0,\label{eq:secondary constraint}
\end{eqnarray}
define the constrained phase space of the model. The symbol ``$\thickapprox$''
means weakly equal, which is used in equality relations only valid
on the constraint surface. There
are two kinds of constraints: first- and second-class ones. A first-class constraint is one whose Poisson brackets, when calculated
in the extended phase space with any other constraint, vanish, whereas a
second class possesses at least one nonvanishing Poisson
bracket with another constraint. In the present case, the Poisson bracket
between the two constraints in Eqs. (\ref{eq:primary constraint})
and (\ref{eq:secondary constraint}) is nonvanishing, so they are
second class.

An important fact about constrained systems is that
the number of propagating degrees of freedom  is different from the
number that one begins with in the Lagrangian.
Given a system described by $N$ degrees of freedom and with $n_{1}$ and $n_{2}$
first- and second-class constraints, respectively, the number of propagating
degrees of freedom is $N-n_{1}-\frac{n_{2}}{2}$. From the above discussion,
the model described by Lagrangian (\ref{eq:Lagrangian})
has $N=4$, $n_{1}=0$, and $n_{2}=2$. So, only three out of the four
degrees of freedom actually propagate in the model. The dynamics of the fields
is governed by the extended Hamiltonian, which is given by the canonical
Hamiltonian $\mathcal{H}_{c}=\Pi^{\mu}B_{\mu}-\mathcal{L}$ up to
additional multiples of the constraints. The coefficients multiplying
the constraints can be determined by consistency requirements in the
case of second-class constraints or remain arbitrary in the case of
first-class ones.

Using the constraints in Eqs. (\ref{eq:primary constraint}) and (\ref{eq:secondary constraint})
and integration by parts, the canonical Hamiltonian can be written as
\begin{equation}
\mathcal{H}=\frac{1}{2}\left(\Pi^{i}\right)^{2}+\frac{1}{4}\left(B_{jk}\right)^{2}-\frac{1}{4}\kappa\left(3B_{0}^{2}+B_{j}^{2}+b^{2}\right)\left(B_{0}^{2}-B_{j}^{2}-b^{2}\right)+B_{i}J^{i}.\label{eq:hamiltonian}
\end{equation}
The situation when $\vec{J}=0$ is the one in which the external matter
fields do no work on the $B_{\mu}$ fields.
A stable model should have a positive Hamiltonian in
this limit. This is not the case for the
Hamiltonian given in Eq. (\ref{eq:hamiltonian}) when general field
configurations are considered. However, positivity can be attained
if the field configurations are restricted to satisfy the condition $\left(B_{0}^{2}-B_{j}^{2}-b^{2}\right)=0$.
It can be shown, using the extended Hamiltonian, that this choice
of the initial conditions is preserved by the field dynamics \cite{stability bumblebee}. This condition
avoids the propagation of the massive mode, and the restricted phase space turns
out to be equivalent to the phase space of the Maxwell electrodynamics in a
nonlinear LV gauge. One of the main goals of this
work is to develop a framework suitable for the quantum analysis of
this stability issue. This will be the subject of the following sections.

\section{Perturbation analysis and Stueckelberg method\label{sec:Perturbation-analysis-and}}

Since the KS model exhibits a phase where Lorentz symmetry
is spontaneously broken, it is convenient to redefine the vector field
$B_{\mu}$ as a perturbation, $\beta_{\mu}$, around its expectation
vacuum value $b_{\mu}$. That is,
\begin{equation}
B_{\mu}=b_{\mu}+\beta_{\mu}.\label{eq:expansion field around vacuum}
\end{equation}
In terms of this expansion, the Lagrangian (\ref{eq:Lagrangian})
is written as
\begin{equation}
\mathcal{L}_{KS}=-\frac{1}{4}\beta_{\mu\nu}\beta^{\mu\nu}-\frac{\kappa}{4}\left(4b_{\mu}\beta^{\mu}b_{\nu}\beta^{\nu}+\beta_{\mu}\beta^{\mu}\beta_{\nu}\beta^{\nu}+4b_{\mu}\beta^{\mu}\beta_{\nu}\beta^{\nu}\right)-\beta_{\mu}J^{\mu}-b_{\mu}J^{\mu},\label{eq:lagrangian beta field}
\end{equation}
where
\begin{equation}
\beta_{\mu\nu}=\partial_{\mu}\beta_{\nu}-\partial_{\nu}\beta_{\mu}.\label{eq:field strength beta}
\end{equation}

The presence of the second-class constraints (\ref{eq:primary constraint})
and (\ref{eq:secondary constraint}) hampers the direct application of the canonical quantization rules. There are
many situations where it is desirable to look for alternatives to the standard Dirac method of quantization of constrained systems.
For gauge-invariant models, which are examples of models presenting first-class
constraints, there are powerful tools associated with this procedure,
like the Gupta-Bleuler and BRST quantization. In this case, the original gauge-invariant Lagrangian is deformed
by adding suitable gauge-noninvariant terms and the Fadeev-Popov
ghosts. The extra degrees of freedom, inserted by the gauge-violating terms, are eliminated by imposing
additional constraints on the final set of quantum states. The absence of unphysical degrees of freedom in the second-class constrained systems foils the direct application of this procedure.

To make use of the same framework described in the quantization of gauge
theories, the model given by Lagrangian (\ref{eq:lagrangian beta field}) will be considered 
as a gauge-fixing limit of some gauge-invariant one. This will be done by enlarging
the field content of the KS model and introducing 
a suitable gauge symmetry, in such a way that the new degrees are of no physical consequence. This technique is known
in the literature as the Stueckelberg method.  Despite the lack of physical consequences, the presence of a new field and a new symmetry provide a greater flexibility in the mathematical treatment of some properties of the model. The successful application of the Stueckelberg procedure in the analysis of the unitarity and renormalizability of massive vector theories is an example of its convenience. In this case, the Proca Lagrangian describes a massive vector field, and there is an apparent incompatibility between power-counting renormalizability and the absence of negative-norm states in the spectrum of the theory. However, by implementing the Stueckelberg method one can define an equivalent Lagrangian where the renormalizability and unitarity are evident \cite{massive vector}. A review of the Stueckelberg method can be found, for example, in Ref. \cite{altaba}.

The Stueckelberg field is introduced in the Lagrangian (\ref{eq:lagrangian beta field}) through the substitution
\begin{equation}
\beta_{\mu}\longrightarrow\beta_{\mu}-\frac{1}{\sqrt{\kappa}}\partial_{\mu}\phi.\label{eq:stueckelberg field}
\end{equation}
With this substitution, the Kosteleck\'{y}-Samuel-Stueckelberg (KSS) Lagrangian is defined as
\begin{equation}
\mathcal{L}_{KSS}=\mathcal{L}_{KS} \left(\beta_{\mu}\longrightarrow\beta_{\mu}-\frac{1}{\sqrt{\kappa}}\partial_{\mu}\phi\right).\label{eq:kss lagrangian}
\end{equation}

This Lagrangian is invariant under the following gauge transformations:
\begin{eqnarray}
\beta_{\mu}^{\prime} & = & \beta_{\mu}+\partial_{\mu}\chi,\label{eq:gauge transf beta}\\
\phi^{\prime} & = & \phi+\sqrt{\kappa}\chi,\label{eq:gauge transf phi}
\end{eqnarray}
where $\chi$ is some arbitrary smooth function of the spacetime coordinates.
The invariance can easily be seen by noticing that the combination on the right-hand side of expression (\ref{eq:stueckelberg field})
is invariant under these transformations. The arbitrariness
in the field content permits one to choose the function $\chi$ in Eqs.
(\ref{eq:gauge transf beta}) and (\ref{eq:gauge transf phi}) such
that the $\phi$ field vanishes, and the original Lagrangian
(\ref{eq:lagrangian beta field}) is recovered. In the standard terminology
of gauge theories, this is known as the unitary gauge. Nevertheless,
more interesting is to take advantage of the gauge freedom of the
Lagrangian (\ref{eq:kss lagrangian}) and make use of
the above mentioned machinery employed in the treatment of gauge-invariant
models. In this vein, the Lagrangian (\ref{eq:kss lagrangian}) will be deformed 
by adding to it a gauge-violating term, promoting, in this way,
the propagation of the gauge degrees of freedom. As a commonly used terminology
in the literature, this term will be referred to as a gauge-fixing term.

It is convenient to choose a gauge-fixing term that provides dynamics for the 0 component
of the $\beta_{\mu}$ field and cancels out the mixing between the
$\beta_{\mu}$ and $\phi$ fields in the Lagrangian (\ref{eq:kss lagrangian}).
The first criterion enables one to avoid the presence of the second-class constraints in Eqs. (\ref{eq:primary constraint}) and (\ref{eq:secondary constraint}),
whereas the second is only for making the future correspondence between
the fields and the particle quantum states more transparent. One can easily verify that the following gauge-fixing Lagrangian
meets these requirements:
\begin{equation}
\mathcal{L}_{gf}=-\frac{1}{2\xi}\left(b^{\nu}b^{\mu}\partial_{\nu}\beta_{\mu}-2\xi\sqrt{\kappa}\phi\right)^{2},\label{eq:gauge fixing Lagrangian}
\end{equation}
where $\xi$ is a parameter whose value can be chosen conveniently.

Gathering this gauge-fixing Lagrangian with $\mathcal{L}_{KSS}$, one gets the total Lagrangian, $\mathcal{L}_{T}=\mathcal{L}_{KSS}+\mathcal{L}_{gf}$, to be discussed from now on. The introduction of a gauge-fixing term into the Lagrangian can sometimes be dangerous, since this term provides dynamics for the unphysical degrees of freedom and can lead to nontrivial consequences. The BRST quantization method is an interesting tool to analyze this issue. In Appendix \ref{appendix a}, it is shown that $\mathcal{L}_{T}=\mathcal{L}_{KSS}+\mathcal{L}_{gf}$ is BRST invariant if the Fadeev-Popov ghosts are added, but these decouple from the other fields and can be discarded without affecting physical results. Furthermore, $\mathcal{L}_{T}$ with the ghost fields included differs from $\mathcal{L}_{KSS}$ by a term that is in the image of the BRST operator. If a physical state is defined  as a state without ghosts, by imposing the Gupta-Bleuler condition, this term does not bring any contribution to the physical states. The net result is that the physical Hilbert space construct from $\mathcal{L}_{T}$ is the same as the one construct from $\mathcal{L}_{KSS}$. For a review of the consequences of the BRST invariance, see for example Ref. \cite{lee}.

The intention of this work is to construct a framework to perform perturbative quantum
calculations with the KS model. As a first effort in this direction, the attention will be
mainly focused on the free part of the total Lagrangian $\mathcal{L}_{KSS}+\mathcal{L}_{gf}$; that is, the current $J^{\mu}$ will be switched
off, and only quadratic terms in the $\beta_{\mu}$ and $\phi$ fields will be considered. By doing this, one assumes that the constant $\kappa$ is sufficiently small to be
considered as a perturbation parameter. Without the interaction terms, the KS Lagrangian (\ref{eq:lagrangian beta field}) turns out to be of the same form as the Proca-like LV theories considered in Refs. \cite{dvali, gabadadze}. In those works, an explicitly LV mass term of the form $m^2 A^{2}_{i}$ is considered along with the Maxwell kinetic term for the vector field $A^{\mu}$ rendering the transverse modes to be massive. In the context of the electron-photon sector of the SME, which is $U(1)$ gauge invariant, there is the possibility that these gauge-violating mass terms can arise as a result of radiative corrections. In the work of Ref. \cite{altschul}, this issue is addressed, and the dispersion relations for a more general class of LV mass terms of the form $M_{\mu\nu}A^{\mu}A^{\nu}$ are discussed. The violation of the Lorentz symmetry in the mentioned works is explicit, since they do not take into account the NG and massive modes emerging from the spontaneous symmetry-breaking mechanism. In the present work, on the other hand, the considered vector field is assumed to describe these NG and massive modes. As a result, despite the form, the quadratic Lagrangian $-\frac{1}{4}\beta_{\mu\nu}\beta^{\mu\nu}-2\kappa b_{\mu}b_{\nu}\beta^{\mu}\beta^{\nu}$ still has a symmetry related to Lorentz invariance of the complete Lagrangian. To verify this, one can perform the infinitesimal transformation $\beta^{\prime}_{\mu}=\beta_{\mu}+\omega_{\mu\nu}b^{\nu}$ with fixed  $b_{\mu}$ in the quadratic Lagrangian and use the antisymmetry of the Lorentz group parameters $\omega_{\mu\nu}$. This nonlinear symmetry is a reminiscence of the Lorentz symmetry present in the full Lagrangian. The Lorentz group acts linearly on the field $B^{\mu}$ via $\Lambda^{\mu}_{\ \ \nu}B^{\nu}$. Since $B_{\mu}=b_{\mu}+ \beta_{\mu}$, an infinitesimal Lorentz transformation $\Lambda^{\mu}_{\ \ \nu}=\delta^{\mu}_{\nu}+\omega^{\mu}_{\ \ \nu}$ acting on $B_{\mu}$ yields $b_{\mu}+\beta^{\prime}_{\mu}=b_{\mu}+\beta_{\mu}+\omega^{\mu}_{\ \ \nu}b^{\nu}+\mathcal{O}(\omega^2)$, where $\omega^{\mu}_{\ \ \nu}$ is supposed to be of the same order of magnitude as the perturbation field $\beta_{\mu}$. This gives the nonlinear transformation of the field $\beta_{\mu}$.  In Ref. \cite{altschul} the mass matrix $M_{\mu\nu}$ cannot assume the form of a product of two vectors, and such a shift symmetry cannot be constructed. For the longitudinal mode, $b\cdot\beta$, this transformation has no effect, but for the transverse mode $\beta^{T}$, which satisfies $b\cdot\beta^{T}=0$, the shift symmetry is expected to avoid the appearance of a mass term generated by quantum corrections.

The stability of the treatment
under the insertion of the self-interactions of the $\beta_{\mu}$
field and with the interactions with the auxiliary field $\phi$, along with the
external matter current, is of great importance, but is beyond the
scope of the present work. Hence, taking into account only quadratic terms in the $\beta_{\mu}$ and $\phi$ fields, one gets the following free Lagrangian:
\begin{equation}
\mathcal{L}_{free}=-\frac{1}{4}\beta_{\mu\nu}\beta^{\mu\nu}-\frac{b^{\nu}b^{\mu}b^{\rho}b^{\sigma}}{2\xi}\partial_{\nu}\beta_{\mu}\partial_{\rho}\beta_{\sigma}-\kappa b_{\mu}b_{\nu}\beta^{\mu}\beta^{\nu}-b_{\mu}b_{\nu}\partial^{\mu}\phi\partial^{\nu}\phi-2\xi\kappa\phi^{2}.\label{eq:total lagrangian}
\end{equation}

The canonical conjugate momenta associated with the fields in this Lagrangian are given by
\begin{eqnarray}
\Pi_{\beta}^{\mu} & = & -\tilde{\eta}^{\mu\sigma}\dot{\beta}_{\sigma}+\Gamma^{\mu\sigma0i}\partial_{i}\beta_{\sigma},\label{eq:beta conjugate momentum}\\
\Pi_{\phi} & = & -2b_{0}b^{\mu}\partial_{\mu}\phi,\label{eq:phi conjugate momentum}
\end{eqnarray}
with
\begin{eqnarray}
\tilde{\eta}^{\mu\sigma} & = & \left(\eta^{00}\eta^{\mu\sigma}-\eta^{\mu0}\eta^{\sigma0}+\frac{b_{0}^{2}b^{\mu}b^{\sigma}}{\xi}\right),\label{eq:eta tilde def}\\
\Gamma^{\mu\sigma0i} & = & \left(\eta^{\mu i}\eta^{0\sigma}-\frac{b_{0}b^{i}b^{\mu}b^{\sigma}}{\xi}\right).\label{eq:gamma def}
\end{eqnarray}
Since the constant vector $b_{\mu}$ is timelike, $\tilde{\eta}^{\mu\sigma}$
is an invertible matrix, and one can invert the relations in Eqs. (\ref{eq:beta conjugate momentum})
and (\ref{eq:phi conjugate momentum}) to write the time derivatives
of the fields in terms of the canonical momenta and the fields themselves.
So, as expected, this is a regular Lagrangian system, and its quantization
follows the standard procedure of considering the observables as quantum
operators acting on the Hilbert space of the particle states and the classical
Poisson brackets being replaced by commutators. The equal-time canonical
commutation relations (ETCR) are, therefore, given by
\begin{eqnarray}
\left[\phi\left(t,\vec{x}\right),\Pi_{\phi}\left(t,\vec{y}\right)\right] & = & i\delta^{3}\left(\vec{x}-\vec{y}\right),\label{eq:phi com rel}\\
\left[\beta_{\nu}\left(t,\vec{x}\right),\Pi_{\beta}^{\mu}\left(t,\vec{y}\right)\right] & = & i\delta_{\nu}^{\mu}\delta^{3}\left(\vec{x}-\vec{y}\right),\label{eq:beta com rel}
\end{eqnarray}
and any other commutator vanishes.

From these commutators and the expressions for the conjugate momenta
(\ref{eq:beta conjugate momentum}) and (\ref{eq:phi conjugate momentum}),
some other useful commutation relations involving fields
and time derivatives of them can be derived. Namely,
\begin{eqnarray}
\left[\phi\left(t,\vec{x}\right),\dot{\phi}\left(t,\vec{y}\right)\right] & = & -\frac{i}{2b_{0}^{2}}\delta^{3}\left(\vec{x}-\vec{y}\right),\label{eq:phi phi dot com rel}\\
\left[\dot{\phi}\left(x\right),\dot{\phi}\left(y\right)\right] & = & \frac{i}{b_{0}^{3}}b^{i}\partial_{i}^{x}\delta^{3}\left(\vec{x}-\vec{y}\right),\\
\left[\dot{\beta}_{\sigma}\left(t,\vec{y}\right),\beta_{\nu}\left(t,\vec{x}\right)\right] & = & i\bar{\eta}_{\sigma\nu}\delta^{3}\left(\vec{x}-\vec{y}\right),\label{eq:beta dot beta com rel}\\
\left[\dot{\beta}_{\mu}\left(t,\vec{x}\right),\dot{\beta}_{\nu}\left(t,\vec{y}\right)\right] & = & -i\bar{\eta}_{\mu\rho}\bar{\eta}_{\nu\sigma}\lambda^{\rho\sigma0i}\partial_{i}^{x}\delta^{3}\left(\vec{x}-\vec{y}\right),\label{eq:beta dot beta dot com rel}
\end{eqnarray}
with $\bar{\eta}$ being the inverse of the $\tilde{\eta}$ matrix
(\ref{eq:eta tilde def}), which is given explicitly by
\begin{equation}
\bar{\eta}_{\mu\nu}=\frac{\eta_{\mu\nu}}{\eta^{00}}+\left(\frac{\xi}{b_{0}^{4}}+\frac{b^{2}}{\eta^{00}b_{0}^{2}}\right)\eta_{0\mu}\eta_{0\nu}-\frac{1}{\eta^{00}b_{0}}\left(\eta_{0\mu}b_{\nu}+\eta_{0\nu}b_{\mu}\right),\label{eq:inverse eta tilde}
\end{equation}
and $\lambda^{\rho\sigma0i}$ is the symmetric combination of the
$\Gamma^{\mu\sigma0i}$'s defined in Eq. (\ref{eq:gamma def}):
\begin{equation}
\lambda^{\rho\sigma0i}=\Gamma^{\rho\sigma0i}+\Gamma^{\sigma\rho0i}.\label{eq:lambda def}
\end{equation}

It can be noticed from the new Lagrangian
(\ref{eq:total lagrangian}) that the number of dynamical degrees of freedom
has increased from three to five as compared with the initial Lagrangian
(\ref{eq:Lagrangian}). The two extra degrees of freedom are related to the $0$
component of the $\beta_{\mu}$ field and to the Stueckelberg field $\phi$, which
came to the fore through the gauge-fixing Lagrangian (\ref{eq:gauge fixing Lagrangian}).
Evidently, if one desires to recover the properties of the initial KS
model, it is necessary to deal properly with these extra degrees of freedom. This will
be done in Sec. \ref{sec:Stability} by choosing a specific region of the full Hilbert space that
accommodates the particle states of the model described by $\mathcal{L}_{free}$,
thwarting the appearance of the extra degrees of freedom in the physical spectrum.

\section{Fourier expansion\label{sec:Fourier-modes-expansion}}

In this section, the relations between the energy and momentum for the particle spectrum of the model described by Lagrangian (\ref{eq:total lagrangian}) are obtained, and the expansion of the fields $\beta_{\mu}$ and $\phi$ in terms of Fourier modes is derived. Obtaining the dispersion relations for the propagating modes and the discussion of their physical properties is the subject of Sec. \ref{disp rel}. Section \ref{part sol} introduces the concept of ``pure-mode solutions,'' which are particular solutions of the equations of motion of the $\beta_{\mu}$ field that satisfy convenient orthogonality relations. Finally, the general solutions for the $\beta_{\mu}$ and $\phi$ fields are derived in Sec. \ref{gen sol}.

\subsection{Dispersion relations \label{disp rel}}

To proceed with the analysis of the model described by Lagrangian
(\ref{eq:total lagrangian}), the equations of motion
are derived. They are
\begin{eqnarray}
\partial_{\mu}\partial^{\mu}\beta^{\nu}-\partial_{\mu}\partial^{\nu}\beta^{\mu}+\frac{b^{\nu}b^{\mu}b^{\rho}b^{\sigma}}{\xi}\partial_{\mu}\partial_{\rho}\beta_{\sigma}-2\kappa b_{\mu}b^{\nu}\beta^{\mu} & = & 0,\label{eq:field eq beta}\\
b^{\mu}b^{\nu}\partial_{\mu}\partial_{\nu}\phi-2\xi\kappa\phi & = & 0.\label{eq:field eq phi}
\end{eqnarray}
Assuming that the fields can be expressed as Fourier integrals, one gets
these equations in momentum space:
\begin{eqnarray}
\left[-p^{2}\eta^{\mu\nu}+p^{\nu}p^{\mu}-\left(\frac{\left(p\cdot b\right)^{2}}{\xi}+2\kappa\right)b^{\mu}b^{\nu}\right]\beta_{\nu}\left(p\right) & = & 0,\label{eq:equation motion beta momentum space}\\
\left(\left(p\cdot b\right)^{2}+2\xi\kappa\right)\phi\left(p\right) & = & 0.\label{eq:equation motion phi momentum space}
\end{eqnarray}
The conditions for the existence of nontrivial solutions for these
equations are given, respectively, by
\begin{eqnarray}
\text{det}\left[-p^{2}\eta^{\mu\nu}+p^{\nu}p^{\mu}-\left(\frac{\left(p\cdot b\right)^{2}}{\xi}+2\kappa\right)b^{\mu}b^{\nu}\right] & = & 0,\label{eq:dispersion rel beta}\\
\left(p\cdot b\right)^{2}+2\kappa\xi & = & 0.\label{eq:dispersion  rel phi}
\end{eqnarray}
The second equation provides the dispersion relation of the particle
associated with the Stueckelberg field, whereas the roots of the first
equation give the dispersion relations for the particles associated
with $\beta_{\mu}$. The latter are promptly obtained by solving
Eq. (\ref{eq:dispersion rel beta}). Since the expression inside the brackets
is a $4\times4$ matrix, an eighth-order polynomial in the momentum
$p$ is expected from Eq. (\ref{eq:dispersion rel beta}), and, in the most general scenario, eight distinct roots. Nevertheless,
the polynomial only presents monomials with even powers in the four-momentum; therefore, it is invariant under the replacement $\left(p^{0},\vec{p}\right)\longrightarrow\left(-p^{0},-\vec{p}\right)$.
The solution with negative energy and negative three-momentum can
be reinterpreted through a parity and time-reversal transformation (PT) as a solution with
positive energy and positive three-momentum. This reflects the fact
that CPT symmetry remains unbroken in this model, as can
be directly seen from the Lagrangian (\ref{eq:total lagrangian}).
The symmetry under charge conjugation is trivial, since the field is real. However, parity and time reversal, considered in  isolation, are not
symmetries of the Lagrangian (\ref{eq:total lagrangian}),
which can also be verified by the noninvariance of Eq. (\ref{eq:dispersion rel beta})
under the replacement $\left(p^{0},\vec{p}\right)\longrightarrow\left(-p^{0},\vec{p}\right)$.

The dispersion relations obtained from Eq. (\ref{eq:dispersion rel beta}) will be labeled by $\lambda=0,1,2,3$.
They are given by 
\begin{alignat}{2}
\lambda&= 0: \quad  &\left(p\cdot b\right)^{2}+2\kappa\xi &= 0,\label{eq:disp rel beta gauge dep}\\
\lambda&= 1,2: \quad  &p^{2} &= 0,\label{eq:disp rel photon}\\
\lambda&= 3: \quad  &\left(p\cdot b\right) &= 0.\label{eq:disp rel tachyon}
\end{alignat}
The reason for the appearance of only three independent dispersion relations,
instead of the four expected from CPT invariance, has to do with the
remaining symmetry after the spontaneous symmetry breaking of Lorentz symmetry
takes place. In the symmetric phase, the degrees of freedom of a four-vector field
can be mapped to those of a spin-$1$ and a spin-$0$ particle.
The presence of a timelike background vector in a CPT invariant field
theory promotes a splitting in the dynamics of these degrees of freedom, and the
particle states organize themselves into classes of opposite spin polarizations.
For the four-vector case, its four degrees of freedom potentially describe
two new spins $0$: the original spin $0$, and the $0$ polarization
of the original spin $1$; and one new spin $1$: the $\pm1$ polarizations
of the original spin $1$. So, these two polarizations of the spin $1$
share the same dispersion relation, and this is the reason for the degeneracy
of the massless pole in Eq. (\ref{eq:disp rel photon}). In the following,
it will be verified that the mode described by Eq. (\ref{eq:disp rel photon})
is indeed a spin-$1$ particle, and it will be identified as the photon.

One can also note the occurrence of the same dispersion relation in
Eqs. (\ref{eq:dispersion  rel phi}) and (\ref{eq:disp rel beta gauge dep}),
as well as their dependence on the gauge-fixing parameter $\xi$.
In fact, these two modes are the ones brought about by the gauge-fixing
Lagrangian (\ref{eq:gauge fixing Lagrangian}) and are, in
this sense, unphysical. Therefore, no concern needs to be dedicated
to their dependence on the gauge-fixing parameter or possible issues
with the appearance of negative energies. However, since the components
of the four-momentum are the reciprocal coordinates of the spacetime
coordinates in the Fourier expansion, they need to be real. This restricts
the gauge parameter to assume only negative values and highlights
a remarkable difference in the role played by the gauge-fixing term
in this model as compared with the SM gauge theories, where
no such dependence of the dispersion relations on the gauge parameter appears, and no restriction in their
values is present.

The four-momentum in the dispersion relation (\ref{eq:disp rel tachyon}) is spacelike and gauge independent.
So, one cannot advocate that the associated excitation will not appear in the physical spectrum.
Indeed, its presence is really an indication of an instability in
the model. This kind of instability should already be expected, since
as seen in Sec. \ref{sec:Spontaneous-Lorentz-symetry}, besides
the NG modes, the KS model propagates a massive mode
that renders the Hamiltonian to be unbounded from below. In that
classical discussion, it was argued that the instability could be avoided
by choosing a suitable slice of the full phase space of the field solutions.
The framework to address this question
in the quantized picture will be discussed in Sec. \ref{sec:Stability}.

\subsection{Suitable particular solutions for the $\beta_{\mu}$ field \label{part sol}}

The decomposition of the vector field $\beta_{\mu}$ in terms of Fourier
modes is subtle. To this end, it is convenient to define first what
will be called ``pure-mode solutions,'' $\beta_{\mu}^{\left(\lambda\right)}\left(\vec{p}\right)$,
that satisfy
\begin{equation}
\left[-p^{2}\eta^{\mu\nu}+p^{\nu}p^{\mu}-\left(\frac{\left(p\cdot b\right)^{2}}{\xi}+2\kappa\right)b^{\mu}b^{\nu}\right]\Bigg|_{p_{0}=p_{0}^{\lambda}\left(\vec{p}\right)}\beta_{\nu}^{\left(\lambda\right)}\left(\vec{p}\right)=0,\label{eq:pure mode solutions}
\end{equation}
where $p_{0}^{\left(\lambda\right)}$ are the solutions for the dispersion
relations (\ref{eq:disp rel beta gauge dep})\textendash(\ref{eq:disp rel tachyon}).
For $\lambda=0,1,2$, there are actually two solutions of the general type
$p_{0\pm}^{\left(\lambda\right)}=f\left(\vec{p}\right)\pm\sqrt{h\left(\vec{p}\right)}$, but they are not independent due to the invariance of the expression inside the brackets under the substitution $p_{\mu}\longrightarrow -p_{\mu}$. So, only one of the two needs to be considered.
Conventionally, it is assumed that $p_{0}^{\left(\lambda\right)}$ corresponds
to $p_{0+}^{\left(\lambda\right)}$. Up to normalization constants,
one can show that these particular solutions are given by
\begin{eqnarray}
\beta_{\mu}^{\left(0\right)}\left(\vec{p}\right) & = & \left(\frac{\vec{b}\cdot\vec{p}}{b_{0}}+\frac{\sqrt{-2\xi\kappa}}{b_{0}},\vec{p}\right)\equiv p_{\mu}^{\left(0\right)}\left(\vec{p}\right),\label{eq:pure mode sol 0}\\
\beta_{\mu}^{\left(i\right)}\left(\vec{p}\right) & = & \epsilon_{\mu}^{\left(i\right)}\left(\vec{p}\right),\ \ i=1,2,\label{eq:pure mode sol 12}\\
\beta_{\mu}^{\left(3\right)}\left(\vec{p}\right) & = & \left(\frac{\vec{b}\cdot\vec{p}}{b_{0}},\vec{p}\right)\equiv p_{\mu}^{\left(3\right)}\left(\vec{p}\right),\label{eq:pure mode sol 3}
\end{eqnarray}
where $\epsilon_{\mu}^{\left(i\right)}\left(\vec{p}\right)$ are two
independent spacelike four-vectors that are simultaneously orthogonal
to $p_{\mu}^{\left(i\right)}=\left(\left|\vec{p}\right|,\vec{p}\right)$
and to the background vector $b_{\mu}$. From their properties, one
can derive the projector on this orthogonal subspace:
\begin{equation}
\sum_{i=}^{2}\epsilon_{\mu}^{\left(i\right)}\left(\vec{p}\right)\epsilon_{\nu}^{\left(i\right)}\left(\vec{p}\right)=-\eta_{\mu\nu}+\frac{1}{\bar{p}\cdot b}\left(b_{\mu}\bar{p}_{\nu}+b_{\nu}\bar{p}_{\mu}\right)-\frac{1}{\left(\bar{p}\cdot b\right)^{2}}\bar{p}_{\mu}\bar{p}_{\nu}.\label{eq:projector orthogonal b p}
\end{equation}
where $\bar{p}_{\mu}\equiv \left(\left|\vec{p}\right|,\vec{p}\right)$.

It can be shown that this projector also appears in the propagator for the $\beta_{\mu}$ field, and, after the exclusion of the unphysical modes from the Fock space, it is identified as the propagator for the transverse physical excitations. Moreover, it also coincides with the propagator for the Maxwell theory in the temporal gauge. There, the background vector, $b_{\mu}$, is assumed to have no physical consequences, since it is introduced only for choosing a particular gauge. However, in the present case, this vector could give rise to measurable effects through the coupling with the matter current, as can be seen from the Lagrangian (\ref{eq:lagrangian beta field}).

Although the solutions (\ref{eq:pure mode sol 0})\textendash(\ref{eq:pure mode sol 3})
are particular ones for the equation of motion in momentum
space (\ref{eq:equation motion beta momentum space}), they
are interesting because they satisfy suitable orthogonal
relations. To derive such relations, Eq. (\ref{eq:pure mode solutions}) is rewritten
as
{\small
\begin{equation}
\left[\left(p_{0}^{\left(\lambda\right)}\right)^{2}\tilde{\eta}^{\mu\nu}-\lambda^{\mu\nu0i}p_{0}^{\left(\lambda\right)}p_{i}\right]\beta_{\nu}^{(\lambda)}\left(\vec{p}\right)=\left[\vec{p}^{2}\eta^{\mu\nu}+\left(\delta_{i}^{\nu}\delta_{j}^{\mu}-\frac{b_{i}b_{j}b^{\mu}b^{\nu}}{\xi}\right)p^{i}p^{j}-2\kappa b^{\mu}b^{\nu}\right]\beta_{\nu}^{\left(\lambda\right)}\left(\vec{p}\right),\label{eq:eq motion expanded}
\end{equation}
}
where $\tilde{\eta}$ and $\lambda^{\mu\nu0i}$, defined respectively in Eqs. (\ref{eq:eta tilde def})
and (\ref{eq:lambda def}), are used. Multiplying both sides of
this equation by $\beta^{\lambda^{\prime}}\left(\vec{p}\right)$,
with $\lambda\neq\lambda^{\prime}$, and subtracting from the analogous
relation with $\lambda$ and $\lambda^{\prime}$ interchanged, yields
\begin{eqnarray}
\beta_{\mu}^{\left(\lambda^{\prime}\right)}\left(\vec{p}\right)\left[\left(p_{0}^{\left(\lambda\right)}+p_{0}^{\left(\lambda^{\prime}\right)}\right)\tilde{\eta}^{\mu\nu}-\lambda^{\mu\nu 0i}p_{i}\right]\beta_{\nu}^{\left(\lambda\right)}\left(\vec{p}\right) & = & 0.
\end{eqnarray}
For general $\lambda$ and $\lambda^{\prime}$, one can write
\begin{equation}
\beta_{\mu}^{\left(\lambda^{\prime}\right)}\left(\vec{p}\right)\left[\left(p_{0}^{\left(\lambda\right)}+p_{0}^{\left(\lambda^{\prime}\right)}\right)\tilde{\eta}^{\mu\nu}-\lambda^{\mu\nu 0i}p_{i}\right]\beta_{\nu}^{\left(\lambda\right)}\left(\vec{p}\right)=\eta^{\lambda\lambda^{\prime}}N^{\left(\lambda\right)}\left(\vec{p}\right).\label{eq:eq:ortho rel pol vec 1}
\end{equation}
When $\lambda^{\prime}=\lambda$, the results need to be calculated explicitly.
Using the explicit results for the pure-mode solutions (\ref{eq:pure mode sol 0})\textendash(\ref{eq:pure mode sol 3}),
one has
\begin{eqnarray}
N^{\left(0\right)} & = & -4b_{0}\kappa\sqrt{-2\kappa\xi},\label{eq:norm factor gauge mode}\\
N^{\left(i\right)} & = & 2\left|\vec{p}\right|,\ \ i=1,2;\label{eq:norm factor photon}\\
N^{\left(3\right)} & = & 0.\label{eq:norm factor tachyon}
\end{eqnarray}

Another useful orthogonality relation can be obtained by multiplying
Eq. (\ref{eq:eq motion expanded}) for $\beta_{\mu}^{\left(\lambda^{\prime}\right)}\left(-\vec{p}\right)$,
switching $\lambda$ for $\lambda^{\prime}$ and $\vec{p}$ by $-\vec{p}$,
and subtracting the obtained expression by the original one multiplied
by $\beta_{\mu}^{\left(\lambda^{\prime}\right)}\left(-\vec{p}\right)$.
This gives
\begin{eqnarray}
\beta_{\mu}^{\left(\lambda\right)}\left(\vec{p}\right)\left[\left(p_{0}^{\left(\lambda^{\prime}\right)}\left(-\vec{p}\right)-p_{0}^{\left(\lambda\right)}\left(\vec{p}\right)\right)\tilde{\eta}^{\mu\nu}+\lambda^{\mu\nu 0i}p_{i}\right]\beta_{\nu}^{\left(\lambda^{\prime}\right)}\left(-\vec{p}\right) & = & 0.\label{eq:eq:ortho rel pol vec 2}
\end{eqnarray}

In Ref. \cite{potting gupta}, the quantization
of the photon sector within the framework of the SME is considered. For the Lagrangian
considered in that work, the general solution of the equations of motion
can be written as a combination of the pure-mode solutions. The
analogous orthogonality relations can be used for writing the creation
and annihilation operators in terms of the fields and canonical momenta
and for getting the algebra of these operators. In the present case, a similar
expansion would run into trouble, since the normalization factor for
the massive mode in Eq. (\ref{eq:norm factor tachyon}) vanishes, and
one cannot invert the expansion for this mode. Furthermore, as was
already emphasized, the expansion in terms of the pure-mode solutions
(\ref{eq:pure mode sol 0})\textendash(\ref{eq:pure mode sol 3})
fails to provide the most general solution of the equation of motion
(\ref{eq:field eq beta}).

\subsection{General solutions \label{gen sol}}

To construct the more general solution for $\beta_{\mu}$ following from the equation of motion  (\ref{eq:field eq beta}),
it is convenient to try to decouple the dynamics for the longitudinal modes. Multiplying
the equation of motion (\ref{eq:field eq beta}) by $\partial_{\mu}$
and $b_{\mu}$ yields the two coupled equations for these longitudinal
modes:
\begin{eqnarray}
\left(b\cdot\partial\right)\partial\cdot\beta-\left(\Box+\frac{b^{2}}{\xi}\left(\left(b\cdot\partial\right)^{2}-2\kappa\xi\right)\right)b\cdot\beta & = & 0,\label{eq:eq motion mom long mode}\\
\left(\left(b\cdot\partial\right)^{2}-2\kappa\xi\right)\left(b\cdot\partial\right)b\cdot\beta & = & 0.\label{eq:eq motion background lon mode}
\end{eqnarray}
The solution for the last equation can be promptly obtained, since it
is completely decoupled from the other modes. In possession of this
solution, one can use it in Eq. (\ref{eq:eq motion mom long mode})
to obtain the solution for $\partial\cdot\beta$. Finally, both solutions can be used in Eq. (\ref{eq:field eq beta}) to get the solution
for the transverse modes.

From Eq. (\ref{eq:eq motion background lon mode}), the solution for $b\cdot\beta$ can be conveniently expressed as
\begin{equation}
b\cdot\beta\left(x\right)=\frac{\xi}{\left(2\pi\right)^{3}}\int d^{4}p\delta\left(\left(\left(p\cdot b\right)^{2}+2\xi\kappa\right)p\cdot b\right)\bar{c}\left(p\right)e^{-ip\cdot x}.\label{eq:fourier decomp beta along b 1}
\end{equation}
$\bar{c}(p)$ is a complex function of the four independent variables $p_0$ and $\vec p$. From the reality of the field $b\cdot\beta$, this function satisfies the condition $\bar{c}^{\dagger}(-p_0,-\vec p)=\bar{c}(p_0,\vec p)$. Using this condition and the properties of the delta function, one obtains
\begin{eqnarray}
b\cdot\beta\left(x\right) & = & \frac{1}{\left(2\pi\right)^{3}|b_{0}|}\int d^{3}p \left(d\left(\vec{p}\right)e^{-ip^{(3)}\cdot x}-\frac{1}{4\kappa}\left(c\left(\vec{p}\right)e^{-ip^{(0)}\cdot x}+c^{\dagger}\left(\vec{p}\right)e^{ip^{(0)}\cdot x}\right)\right) \label{eq:fourier decomp beta along b}
\end{eqnarray}
where
\begin{equation}
c\left(\vec{p}\right)\equiv \bar{c}\left(\frac{\vec{b}\cdot\vec{p}}{b_{0}}+\frac{\sqrt{-2\xi\kappa}}{b_{0}},\vec{p}\right)\label{eq:def c of p}
\end{equation}
and
\begin{equation}
d\left(\vec{p}\right)\equiv-\frac{1}{2\kappa}\bar{c}\left(\frac{\vec{b}\cdot\vec{p}}{b_{0}},\vec{p}\right)=d^{\dagger}\left(-\vec{p}\right).\label{eq:def d of p}
\end{equation}

The solution for $\partial\cdot\beta$ in Eq. (\ref{eq:eq motion mom long mode})
can be constructed as the sum of the solution of the homogeneous equation
$\left(b\cdot\partial\right)\partial\cdot\beta=0$ plus a particular
solution for the inhomogeneous one, since the inhomogeneous part is
explicitly known from Eq. (\ref{eq:fourier decomp beta along b}).
The solution for the homogeneous equation can be derived straightforwardly
following the previous reasoning to reach the solution for the $b\cdot\beta$ field in Eq. (\ref{eq:fourier decomp beta along b}). Denoting $\partial\cdot\beta\left(x\right)$
as $S\left(x\right)$, one has
\begin{equation}
S^{H}\left(x\right)=\frac{1}{\left(2\pi\right)^{3}|b_{0}|}\int d^{3}ps\left(\vec{p}\right)e^{-ip^{(3)}\cdot x},\label{eq:div beta homogeneous}
\end{equation}
where $s^{\dagger}\left(-\vec{p}\right)=s\left(\vec{p}\right)$, and
the superscript $H$ on the left-hand side of this equation stands for homogeneous.

For the particular solution $S^{P}$, one could make use of the Green function method. The caveat here is that the convolution of the Green
function for the operator $b\cdot\partial$ with the distribution
$\delta\left(\left(\left(p\cdot b\right)^{2}+2\xi\kappa\right)p\cdot b\right)$
is ill defined. Here, only the final solution for $S^{P}$ is presented,
leaving the details for Appendix \ref{appendix b}:
\begin{eqnarray}
S^{P}\left(x\right) & = & -i\Box\int\frac{d^{3}p}{4\left|b_{0}\right|\sqrt{-2\kappa\xi}}\left(c\left(\vec{p}\right)e^{-ip^{\left(0\right)}\cdot x}-c^{\dagger}\left(\vec{p}\right)e^{ip^{\left(0\right)}\cdot x}\right)\nonumber \\
 &  & +\left(\Box-2\kappa b^{2}\right)\int\frac{d^{3}p}{b_{0}^{2}}x^{0}d\left(\vec{p}\right)e^{-ip^{\left(3\right)}\cdot x}.\label{eq:div beta particular}
\end{eqnarray}
Finally, this solution and the homogeneous part from Eq.
(\ref{eq:div beta homogeneous}) can be used, along with the solution for $b\cdot\beta$
in Eq. (\ref{eq:fourier decomp beta along b}), to obtain the inhomogeneous
part of the differential equation (\ref{eq:field eq beta}).
The solution for the entire field $\beta_{\mu}$ can again be expressed
as a sum of a homogeneous plus an inhomogeneous part. The homogeneous
part can be written as
\begin{equation}
\beta_{\mu}^{H}\left(x\right)=\frac{1}{\left(2\pi\right)^{3}}\int d^{4}pa_{\mu}\left(p\right)\delta\left(p^{2}\right)e^{-ip\cdot x},\label{eq:beta homogeneous}
\end{equation}
where the set of the four vectors $a_{\mu}\left(p\right)\delta\left(p^{2}\right)$,
defined for each momentum $\vec{p}$, can be expanded in terms of
some convenient complete basis of vectors for each point $\vec{p}$.
For the present purposes, a suitable basis can be built using the background
vector $b_{\mu}$, the lightlike four-momentum $\bar{p}_{\mu}\equiv\left(\left|\vec{p}\right|,\vec{p}\right)$,
and the two spacelike vectors in Eq. (\ref{eq:pure mode sol 12}).
Since $b_{\mu}$ is timelike and the $\epsilon_{\mu}^{\left(i\right)}\left(\vec{p}\right)$'s
are simultaneously orthogonal to $p_{\mu}$ and $b_{\mu}$, this
set of vectors forms a complete basis for each momentum $\vec{p}$.
In terms of this basis, the set of vectors $a_{\mu}\left(\vec{p}\right)\equiv a_{\mu}\left(\left|\vec{p}\right|,\vec{p}\right)$
can be expressed as
\begin{equation}
a_{\mu}\left(\vec{p}\right)=\sum_{i=1}^{2}a^{\left(i\right)}\left(\vec{p}\right)\epsilon_{\mu}^{\left(i\right)}\left(\vec{p}\right)+a^{\left(3\right)}\left(\vec{p}\right)\bar{p}_{\mu}+a^{\left(4\right)}\left(\vec{p}\right)b_{\mu}.\label{eq:expansion vector field in a complete basis}
\end{equation}

Concerning the particular solution, there is no obstruction for the
convolution of the Green function of the d'Alembertian operator with
the expressions in Eqs. (\ref{eq:fourier decomp beta along b}), (\ref{eq:div beta homogeneous}),
and (\ref{eq:div beta particular}). Referring to the Green function
of the $\Box$ operator as $G^{\left(1\right)}$, one can write formally
\begin{eqnarray}
\beta_{\mu}^{P}\left(x\right) & = & \partial_{\mu}\left(G^{\left(1\right)}\ast S^{H}\right)+\Box\partial_{\mu}\left(G^{\left(1\right)}\ast G^{\left(3\right)}\ast C\right)+2\kappa b_{\mu}\left(G^{\left(1\right)}\ast D\right)\nonumber \\
 &  & -\left(\Box-2\kappa b^{2}\right)\partial_{\mu}\frac{d}{d\tau}\left(G^{\left(1\right)}\ast D\left(x;\tau\right)\right)\Big|_{\tau=0},\label{eq:beta part sol}
\end{eqnarray}
where the symbol ``$\ast$'' means convolution, and the functions
$C\left(x\right)$ and $D\left(x\right)$ are defined in Eqs. (\ref{eq:def function C}) and (\ref{eq:def function D}),
respectively. $G^{\left(3\right)}$ is the Green function for the
operator $b\cdot\partial$, given in Eq. (\ref{eq:green function 3}),
and the Green function $G^{\left(1\right)}$ can be chosen to be
\begin{equation}
G^{\left(1\right)}=-\frac{1}{\left(2\pi\right)^{4}}\int d^{4}p\frac{e^{-ip\cdot x}}{p^{2}+i\epsilon}.
\end{equation}

Since the equation of motion (\ref{eq:field eq beta}) for the $\beta_{\mu}$ field is
second order in time, one should expect the presence of four pairs
of arbitrary functions of the three-momenta to be fixed by the initial
conditions. However, it can be seen from Eqs. (\ref{eq:fourier decomp beta along b}),
(\ref{eq:div beta homogeneous}), (\ref{eq:div beta particular}),
and (\ref{eq:expansion vector field in a complete basis}) that there are
six pairs at disposal instead: $\left(a^{\left(l\right)}\left(\vec{p}\right),a^{\dagger\left(l\right)}\left(\vec{p}\right)\right)$
with $l=1,\dots,4$; $\left(c\left(\vec{p}\right),c^{\dagger}\left(\vec{p}\right)\right)$;
and $\left(s\left(\vec{p}\right),d\left(\vec{p}\right)\right)$. This
apparent overcounting problem is solved when it is imposed that $b\cdot\beta$
and $\partial\cdot\beta$, calculated from the Eqs. (\ref{eq:beta homogeneous})
and (\ref{eq:beta part sol}), match the expressions in Eqs. (\ref{eq:fourier decomp beta along b}),
(\ref{eq:div beta homogeneous}), and (\ref{eq:div beta particular}).
These two conditions imply in the vanishing of the functions $a^{\left(3\right)}$
and $a^{\left(4\right)}$ in the expansion (\ref{eq:expansion vector field in a complete basis}).
Despite the length, for convenience, the final result
for the expansion of the $\beta_{\mu}$ field is presented here:
\begin{eqnarray}
\beta_{\mu}\left(x\right) & = & \frac{1}{\left(2\pi\right)^{3}}\int\frac{d^{3}p}{2\left|\vec{p}\right|}\sum_{i=1}^{2}\left(a^{\left(i\right)}\left(\vec{p}\right)\epsilon_{\mu}^{\left(i\right)}\left(\vec{p}\right)e^{-ip^{\left(1\right)}\cdot x}+a^{\dagger\left(i\right)}\left(\vec{p}\right)\epsilon_{\mu}^{\left(i\right)}\left(\vec{p}\right)e^{ip^{\left(1\right)}\cdot x}\right)\nonumber \\
 &  & -\frac{i}{4\kappa |b_{0}|\sqrt{-2\kappa\xi}}\partial_{\mu}\int\frac{d^{3}p}{\left(2\pi\right)^{3}}\left(c\left(\vec{p}\right)e^{-ip^{\left(0\right)}\cdot x}-c^{\dagger}\left(\vec{p}\right)e^{ip^{\left(0\right)}\cdot x}\right)\nonumber \\
 &  & +\frac{2\kappa b^{2}}{\left(2\pi\right)^{3}b_{0}^{2}}\int d^{3}p\left(\frac{\delta_{\mu}^{0}-ix^{0}p_{\mu}^{\left(3\right)}}{\left(p^{\left(3\right)}\right)^{2}}-\frac{2p_{0}^{\left(3\right)}p_{\mu}^{\left(3\right)}}{\left(p^{\left(3\right)}\right)^{4}}\right)d\left(\vec{p}\right)e^{-ip^{\left(3\right)}\left(\vec{p}\right)\cdot x}\nonumber \\
 &  & -\frac{2\kappa b_{\mu}}{\left(2\pi\right)^{3}\left|b_{0}\right|}\int d^{3}p\frac{e^{-ip^{\left(3\right)}\cdot x}}{\left(p^{\left(3\right)}\right)^{2}}d\left(\vec{p}\right)-\frac{1}{\left(2\pi\right)^{3}\left|b_{0}\right|}\partial_{\mu}\int d^{3}p\frac{e^{-ip^{\left(3\right)}\cdot x}}{\left(p^{\left(3\right)}\right)^{2}}s\left(\vec{p}\right)\nonumber \\
 &  & +\partial_{\mu}\int\frac{d^{3}p}{\left(2\pi\right)^{3}b_{0}^{2}}x^{0}d\left(\vec{p}\right)e^{-ip^{\left(3\right)}\cdot x}.\label{eq:expansion complete field beta}
\end{eqnarray}

Concerning the general solution for the Stueckelberg field $\phi\left(x\right)$, one can notice from Eq. (\ref{eq:equation motion phi momentum space}) that its solution can be expressed as
\begin{equation}
\phi\left(x\right)=\frac{1}{\left(2\pi\right)^{3}2\sqrt{\kappa}}\int d^{4}p\delta\left(\left(p\cdot b\right)^{2}+2\xi\kappa\right)\bar{g}\left(p\right)e^{-ip\cdot x}.\label{eq:sol phi field}
\end{equation}
Using the properties of the delta function, this expansion yields
\begin{equation}
\phi\left(x\right)=\frac{1}{\left(2\pi\right)^{3}}\int\frac{d^{3}p}{4\sqrt{\kappa}|b_{0}|\sqrt{-2\xi\kappa}}
\left(g\left(\vec{p}\right)e^{-ip^{(0)}\cdot x}+g^{\dagger}\left(\vec{p}\right)e^{ip^{(0)}\cdot x}\right),\label{eq:sol phi field 2}
\end{equation}
since $\phi$ is a real field, and therefore
$\bar{g}^{\dagger}\left(-p_{0},-\vec{p}\right)=\bar{g}\left(p_{0},\vec{p}\right)$.
The function $g\left(\vec{p}\right)$ was also defined by  $\bar{g}\left(\frac{\vec{b}\cdot\vec{p}}{b_{0}}+\frac{\sqrt{-2\xi\kappa}}{b_{0}},\vec{p}\right)$.

The operators $g\left(\vec{p}\right)$ and $g^{\dagger}\left(\vec{p}\right)$
will soon be identified as the annihilation and creation operators of the Stueckelberg field. The momentum-space operators for the transverse field and for the gauge
modes in expansion (\ref{eq:expansion complete field beta}) arrange themselves in the standard way, and it will be verified in
the next section that they can be indeed identified as creation
and annihilation operators for these modes. Nevertheless, the expansion
for the longitudinal sector is not so enlightening, and the role of the operators $s\left(\vec{p}\right)$ and $d\left(\vec{p}\right)$
in the structure of the Fock space is unclear. This issue will now be addressed.

\section{Stability\label{sec:Stability}}

In this section, the conditions for the suppression of the unphysical modes are discussed and implemented. In Sec. \ref{fourier algebra}, the commutation relations for the Fourier modes that appear in field expansions (\ref{eq:expansion complete field beta}) and (\ref{eq:sol phi field 2}) are obtained, and their main properties are analyzed. The Gupta-Bleuler condition for the absence of the gauge modes is also discussed in this subsection. In Sec. \ref{tachyon absence}, the two Fourier-mode operators for the longitudinal sector, $s$ and $d$, are mapped into new ones with a simpler action on the Fock space, and a condition for the absence of tachyons is obtained from them.

\subsection{Fourier-Mode algebra \label{fourier algebra}}

The Fourier expansions for the free fields $\phi$ and $\beta_{\mu}$
in Eqs. (\ref{eq:expansion complete field beta}) and (\ref{eq:sol phi field 2})
are very convenient to make perturbative calculations. However,
one still needs to address the question of the fate of the gauge and
Stueckelberg modes, which are unphysical, and the more subtle question
of the presence of tachyonic excitation in the spectrum of the model.
For this intent, the Fock space of the
present model will be constructed, and the implementation of the conditions to handle 
properly the existence of the unphysical excitations will be discussed in this section.

To begin with, the identification of the right operators that create
and annihilate all the propagating modes in this theory is useful. So, one proceeds
with the inversion of the expansions (\ref{eq:expansion complete field beta}) and (\ref{eq:sol phi field 2}). It can be directly shown that the modes $g$ and $g^{\dagger}$ can be expressed in terms of $\phi$ and $\dot{\phi}$ as
\begin{eqnarray}
g\left(\vec{p}\right) & = & \int d^{3}xe^{ip^{\left(0\right)}\cdot x}\left(i\overleftrightarrow{\partial_{0}^{S}}2\sqrt{\kappa}b_{0}^{2}+4\sqrt{\kappa}b_{0}\sqrt{-2\kappa\xi}\right)\phi\left(x\right),\label{eq:g of p in terms of phi and phi dot}\\
g^{\dagger}\left(\vec{p}\right) & = & \int d^{3}xe^{-ip^{\left(0\right)}\cdot x}\left(-i\overleftrightarrow{\partial_{0}^{S}}2\sqrt{\kappa}b_{0}^{2}+4\sqrt{\kappa}b_{0}\sqrt{-2\kappa\xi}\right)\phi\left(x\right),\label{eq:g dagger of p in terms of phi and phi dot}
\end{eqnarray}
with $x^{0}$ arbitrary, and for two arbitrary functions, $f_{1}$ and
$f_{2}$, $\overleftrightarrow{\partial_{0}^{S}}$ is defined by
\begin{equation}
f_{1}\left(x\right)\overleftrightarrow{\partial_{0}^{S}}f_{2}\left(x\right)\equiv f_{1}\left(x\right)\partial_{0}f_{2}\left(x\right)+\partial_{0}f_{1}\left(x\right)f_{2}\left(x\right),
\end{equation}
where ``$S$"' stands for symmetric to differ from the antisymmetric, $\overleftrightarrow{\partial_{0}^{A}}$, defined by

\begin{equation}
f_{1}\left(x\right)\overleftrightarrow{\partial_{0}^{A}}f_{2}\left(x\right)\equiv f_{1}\left(x\right)\partial_{0}f_{2}\left(x\right)-\partial_{0}f_{1}\left(x\right)f_{2}\left(x\right).
\end{equation}

If the only vectors that appeared in the expansion (\ref{eq:expansion complete field beta}) for the $\beta_{\mu}$ field were the pure-mode solutions
 (\ref{eq:pure mode sol 0})\textendash(\ref{eq:pure mode sol 3}),
one could obtain the inverse of that expansion using the orthogonality relations (\ref{eq:eq:ortho rel pol vec 1})
and (\ref{eq:eq:ortho rel pol vec 2}). This would give
\begin{eqnarray}
a^{\left(i\right)}\left(\vec{p}\right) & = & -\int d^{3}xe^{ip^{\left(1\right)}\cdot x}\left(i\overleftrightarrow{\partial_{0}^{A}}\tilde{\eta}^{\mu\nu}-\lambda^{\mu\nu0i}p_{i}\right)\epsilon_{\mu}^{(i)}\left(\vec{p}\right)\beta_{\nu}\left(x\right),\label{eq:anihilation transverse operator}\\
a^{\dagger\left(i\right)}\left(\vec{p}\right) & = & -\int d^{3}xe^{-ip^{\left(1\right)}\cdot x}\left(-i\overleftrightarrow{\partial_{0}^{A}}\tilde{\eta}^{\mu\nu}-\lambda^{\mu\nu0i}p_{i}\right)\epsilon_{\mu}^{(i)}\left(\vec{p}\right)\beta_{\nu}\left(x\right),\label{eq:creation transverse operator}\\
c\left(\vec{p}\right) & = & \int d^{3}xe^{ip^{\left(0\right)}\cdot x}\left(i\overleftrightarrow{\partial_{0}^{A}}\tilde{\eta}^{\mu\nu}-\lambda^{\mu\nu0i}p_{i}\right)p_{\mu}^{\left(0\right)}\left(\vec{p}\right)\beta_{\nu}\left(x\right),\label{eq:anihilation gauge mode operator}\\
c^{\dagger}\left(\vec{p}\right) & = & \int d^{3}xe^{-ip^{\left(0\right)}\cdot x}\left(-i\overleftrightarrow{\partial_{0}^A}\tilde{\eta}^{\mu\nu}-\lambda^{\mu\nu0i}p_{i}\right)p_{\mu}^{\left(0\right)}\left(\vec{p}\right)\beta_{\nu}\left(x\right).\label{eq:creation gauge mode operator}
\end{eqnarray}
However, besides the pure-mode solutions, there are other
four-vectors composing the full expansion (\ref{eq:expansion complete field beta}). Thence,
extra algebraic relations between the vector quantities
that appear in this expansion would be needed. It turns out that these extra contributions,
coming from the substitution of the entire field $\beta_{\mu}$ in
the expressions above, cancel out, and Eqs. (\ref{eq:anihilation transverse operator})\textendash(\ref{eq:creation gauge mode operator})
are in fact the right relations for these modes.

The last two operators, $s\left(\vec{p}\right)$ and $d\left(\vec{p}\right)$, can be obtained
by considering $\partial\cdot\beta$ and \mbox{$b\cdot\beta$}, respectively,
directly in the expansion (\ref{eq:expansion complete field beta}) for $\beta_{\mu}$ 
and performing suitable Fourier transformations on the result. The
outcome of this procedure is given by
\begin{eqnarray}
d\left(\vec{p}\right) & = & \left|b_{0}\right|\int d^{3}xe^{ip^{\left(3\right)}\cdot x}b\cdot\beta\left(x\right)+\frac{1}{4\kappa}\left(c\left(\vec{p}\right)e^{-i\frac{\sqrt{-2\kappa\xi}}{b_{0}}x^{0}}+c^{\dagger}\left(-\vec{p}\right)e^{i\frac{\sqrt{-2\kappa\xi}}{b_{0}}x^{0}}\right),\label{eq:d operator}\\
s\left(\vec{p}\right) & = & -i\frac{1}{4\kappa\sqrt{-2\kappa\xi}}\left(\left(p^{\left(0\right)}\left(\vec{p}\right)\right)^{2}c\left(\vec{p}\right)e^{-i\frac{\sqrt{-2\kappa\xi}}{b_{0}}x^{0}}-\left(p^{\left(0\right)}\left(-\vec{p}\right)\right)^{2}c^{\dagger}\left(-\vec{p}\right)e^{i\frac{\sqrt{-2\kappa\xi}}{b_{0}}x^{0}}\right)\nonumber \\
 &  & +\frac{1}{|b_{0}|}d\left(\vec{q}\right)\left(2ip_{0}^{\left(3\right)}\left(\vec{p}\right)+x^{0}\left(\left(p^{\left(3\right)}\left(\vec{p}\right)\right)^{2}+2\kappa b^{2}\right)\right)+|b_{0}|\int d^{3}xe^{ip^{\left(3\right)}\cdot x}\partial\cdot\beta\left(x\right).\label{eq:s operator}
\end{eqnarray}
If the expressions (\ref{eq:anihilation gauge mode operator}) and (\ref{eq:creation gauge mode operator}) for $c\left(\vec{p}\right)$ and $c^{\dagger}\left(\vec{p}\right)$, respectively,
are used in the equations for $d\left(\vec{p}\right)$ and $s\left(\vec{p}\right)$
above, this completes the task of writing the Fourier-mode operators in
terms of the fields and their time derivatives.

From the expressions (\ref{eq:g of p in terms of phi and phi dot}),
(\ref{eq:g dagger of p in terms of phi and phi dot}), (\ref{eq:anihilation gauge mode operator}), and (\ref{eq:creation gauge mode operator}) for $g$, $g^{\dagger}$, $c$, and $c^{\dagger}$, respectively,
and using the ETCR in Eqs. (\ref{eq:phi com rel})\textendash(\ref{eq:beta dot beta dot com rel})
along with the orthogonality relations in Eqs. (\ref{eq:eq:ortho rel pol vec 1})
and (\ref{eq:eq:ortho rel pol vec 2}), one can get the algebra of the Fourier
modes
\begin{eqnarray}
\left[g\left(\vec{p}\right),g^{\dagger}\left(\vec{q}\right)\right] & = & -\left(2\pi\right)^{3}4|b_{0}|\kappa\sqrt{-2\xi\kappa}\delta\left(\vec{p}-\vec{q}\right),\label{eq:g algebra}\\
\left[c\left(\vec{p}\right),c^{\dagger}\left(\vec{q}\right)\right] & = & \left(2\pi\right)^{3}4|b_{0}|\kappa\sqrt{-2\kappa\xi}\delta\left(\vec{p}-\vec{q}\right),\label{eq:c algebra}\\
\left[a^{(i)}\left(\vec{p}\right),a^{\dagger (j)}\left(\vec{q}\right)\right] & = & -\left(2\pi\right)^{3}\delta^{ij}2\left|\vec{p}\right|\delta\left(\vec{p}-\vec{q}\right),\label{eq:a algebra}\\
\left[s\left(\vec{p}\right),s\left(\vec{q}\right)\right] & = & -\left(2\pi\right)^{3}\delta\left(\vec{p}+\vec{q}\right)2b^{2}p_{0}^{\left(3\right)}\left(\vec{p}\right),\label{eq:s algebra}\\
\left[s\left(\vec{p}\right),d\left(\vec{q}\right)\right] & = & \frac{i}{2\kappa}\left(2\pi\right)^{3}|b_{0}|\left(q^{\left(3\right)}\right)^{2}\delta\left(\vec{p}+\vec{q}\right).\label{eq:s d algebra}
\end{eqnarray}
All the other possible commutators vanish.

The algebraic properties of the operators $\left(g\left(\vec{p}\right),g^{\dagger}\left(\vec{q}\right)\right)$,
$\left(c\left(\vec{p}\right),c^{\dagger}\left(\vec{q}\right)\right)$,
and $\left(a^{(i)}\left(\vec{p}\right),a^{\dagger (i)}\left(\vec{q}\right)\right)$
are identical to the standard algebra of annihilation and creation
operators. In this way, the vacuum of the theory can be defined as the
state annihilated by the operators $g\left(\vec{p}\right)$, $c\left(\vec{p}\right)$,
and $a\left(\vec{p}\right)$. However, due to the plus sign in the
commutator between $c\left(\vec{p}\right)$ and $c^{\dagger}\left(\vec{q}\right)$,
the Fock space generated from the vacuum state by
the successive operation of the creation operators $g^{\dagger}\left(\vec{q}\right)$,
$c^{\dagger}\left(\vec{q}\right)$, and $a^{\dagger}\left(\vec{q}\right)$
can present negative-norm states and cannot correspond to the
physical Hilbert space. Despite the right sign in the commutator
between $g\left(\vec{p}\right)$ and $g^{\dagger}\left(\vec{q}\right)$,
it depends on the gauge parameter as well as $\left[c\left(\vec{p}\right),c^{\dagger}\left(\vec{q}\right)\right]$.
These modes were both introduced through the gauge-fixing term $\left(b^{\nu}b^{\mu}\partial_{\nu}\beta_{\mu}-2\xi\sqrt{\kappa}\phi\right)$,
and for their elimination, one imposes that the expectation value of this term between physical states must vanish. Moreover, 
to preserve the linear structure of the Hilbert space, one follows the Gupta-Bleuler procedure by imposing that the physical states must belong to the kernel of the annihilation part of the gauge-fixing term. That is,
\begin{equation}
\left(b\cdot\partial b\cdot\beta-2\xi\sqrt{\kappa}\phi\right)^{+}\left|\text{Phys}\right\rangle =0.\label{eq:gupta-bleuler condition}
\end{equation}

The expansion (\ref{eq:fourier decomp beta along b}) for the $b\cdot\beta$ field 
contains, besides the modes $c\left(\vec{p}\right)$ and $c^{\dagger}\left(\vec{p}\right)$,
the Fourier mode $d\left(\vec{p}\right)$, and the condition (\ref{eq:gupta-bleuler condition}) would also contain gauge-independent modes. It turns out that the operator $b\cdot\partial$
acting on $b\cdot\beta$ kills exactly the contribution for the $d\left(\vec{p}\right)$
mode, and the condition (\ref{eq:gupta-bleuler condition}) only
contains the annihilation operators $c\left(\vec{p}\right)$ and $g\left(\vec{p}\right)$
indeed. Using the expansions (\ref{eq:fourier decomp beta along b}) and (\ref{eq:sol phi field 2}) for $b\cdot\beta$ and $\phi$, respectively, the condition (\ref{eq:gupta-bleuler condition})
is equivalent to
\begin{equation}
\left(ic\left(\vec{p}\right)+g\left(\vec{p}\right)\right)\left|\text{Phys}\right\rangle =0 \label{absence ghost}.
\end{equation}

The algebraic properties of the $s\left(\vec{p}\right)$ and $d\left(\vec{p}\right)$ operators in Eqs. (\ref{eq:s algebra}) and (\ref{eq:s d algebra}) require further analysis. The appearance of the unusual $i$ factor on the right-hand side of Eq. (\ref{eq:s d algebra}) is consistent with the constraint conditions obeyed by these operators: $s^{\dagger}\left(-\vec{p}\right)=s\left(\vec{p}\right)$ and $d^{\dagger}\left(-\vec{p}\right)=d\left(\vec{p}\right)$. One can verify this fact by taking the adjoint of Eq. (\ref{eq:s d algebra}) and making the replacements $\vec{p}\longrightarrow-\vec{p}$ and $\vec{q}\longrightarrow-\vec{q}$. Furthermore, these operators cannot be interpreted directly as creation and annihilation operators without contradicting their algebraic relations.
As a single example, suppose that $s\left(\vec{p}\right)$ is
an annihilation operator. Therefore, it annihilates the vacuum, and
$\left\langle 0\left|\left[s\left(\vec{p}\right),s\left(\vec{q}\right)\right]\right|0\right\rangle $
should also give a null result, but this contradicts the nonvanishing right-hand side
of Eq. (\ref{eq:s algebra}). Many other examples can be given whenever
$s\left(\vec{p}\right)$ or $d\left(\vec{p}\right)$ are supposed to
be creation or annihilation operators. The correct interpretation of the
role of these operators in the structure of the Fock space plays a fundamental role
in the present analysis, since this sector houses the massive tachyonic mode,
and the stability of the model demands its suppression. To accomplish this purpose, it is convenient to construct other operators from $s$ and $d$ such that they have a simpler action in the Fock space.

\subsection{Condition for the absence of tachyons \label{tachyon absence}}

Due to the constraints obeyed by $s\left(\vec{p}\right)$ and $d\left(\vec{p}\right)$, they carry enough information to be mapped in a one-to-one way into two adjointed-related complex operators.  Let the latter be denoted by $\tau(\vec{p})$ and $\tau^{\dagger}(\vec{p})$. Consider the complex linear transformation
\begin{equation}
\left(\begin{array}{c}
\tau\left(\vec{p}\right)\\
\tau^{\dagger}\left(-\vec{p}\right)
\end{array}\right)=\left(\begin{array}{cc}
\rho\left(\vec{p}\right) & \sigma\left(\vec{p}\right)\\
\rho^{\ast}\left(-\vec{p}\right) & \sigma^{\ast}\left(-\vec{p}\right)
\end{array}\right)\left(\begin{array}{c}
s\left(\vec{p}\right)\\
d\left(\vec{p}\right)
\end{array}\right), \label{eq:tau tau dagger s d relation}
\end{equation}
where $\rho$ and $\sigma$ are two arbitrary complex functions of $\vec{p}$, such that $\tau$ and $\tau^{\dagger}$ satisfy the following commutation relations:
\begin{eqnarray}
\left[\tau\left(\vec{p}\right),\tau\left(\vec{q}\right)\right] & = & 0,\label{eq:tau tau algebra}\\
\left[\tau\left(\vec{p}\right),\tau^{\dagger}\left(\vec{q}\right)\right] & \neq & 0. \label{eq:tau tau dagger algebra}
\end{eqnarray}

These conditions, together with the requirement that the transformation (\ref{eq:tau tau dagger s d relation}) be one to one, yield
\begin{eqnarray}
\left(\rho\left(\vec{p}\right)\rho\left(-\vec{p}\right)\left(-2b^{2}p_{0}^{\left(3\right)}\left(\vec{p}\right)\right)+\left(\rho\left(\vec{p}\right)\sigma\left(-\vec{p}\right)-\rho\left(-\vec{p}\right)\sigma\left(\vec{p}\right)\right)\frac{i}{2\kappa}b_{0}\left(p^{\left(3\right)}\right)^{2}\right) & = & 0,\label{eq:rho sigma constraint 1}\\
\left(\left|\rho\left(\vec{p}\right)\right|^{2}\left(-2b^{2}p_{0}^{\left(3\right)}\left(\vec{p}\right)\right)+\left(\rho\left(\vec{p}\right)\sigma^{\ast}\left(\vec{p}\right)-\rho^{\ast}\left(\vec{p}\right)\sigma\left(\vec{p}\right)\right)\frac{i}{2\kappa}b_{0}\left(p^{\left(3\right)}\right)^{2}\right) & \neq & 0,\\
\rho\left(\vec{p}\right)\sigma^{\ast}\left(-\vec{p}\right)-\rho^{\ast}\left(-\vec{p}\right)\sigma\left(\vec{p}\right) & \neq & 0.\label{eq:rho sigma constraint 3}
\end{eqnarray}

It is straightforward to show that there are numerous possibilities
for the choices of $\rho$ and $\sigma$ that satisfy these conditions.
It seems that no specific choice is preferable to any other. For the present purposes, it is just assumed that some choice
of $\rho$ and $\sigma$ is made such that it satisfies the requirements
(\ref{eq:rho sigma constraint 1})\textendash(\ref{eq:rho sigma constraint 3}).
Now, since the algebra in Eqs. (\ref{eq:tau tau algebra}) and (\ref{eq:tau tau dagger algebra})
is the standard one for creation and annihilation operators, $\tau^{\dagger}$ and $\tau$ can be identified as
creation and annihilation operators for the tachyonic
mode, respectively. Thus, as a last condition in the definition of a physical
state, one imposes
\begin{equation}
\tau\left(\vec{p}\right)\left|\text{Phys}\right\rangle =0.\label{eq:absence of tackyon condition}
\end{equation}
This condition together with the condition (\ref{absence ghost}) for the suppression of the gauge modes is sufficient to show that the only contribution for the expansion of the $\beta_{\mu}$ and $\phi$ fields between physical states comes from the transverse modes. These, in turn, are creation and annihilation operators for a massless spin-1 particle, which can be seen from the expression for the transverse projector (\ref{eq:projector orthogonal b p}) and the algebraic relation (\ref{eq:a algebra}). As already mentioned before, the projection operator (\ref{eq:projector orthogonal b p}) is the effective physical propagator at the tree level for the $\beta_{\mu}$ field, and it coincides with the photon propagator of the Maxwell theory in the temporal gauge. In this way, the KS model provides an alternative to the gauge-invariant description of the photon by considering it an as NG mode arising from the spontaneous breaking of Lorentz symmetry.

Another way to get the condition (\ref{eq:absence of tackyon condition}) is by imposing the positiveness of
the expectation value between physical states of the Hamiltonian
associated with the Lagrangian (\ref{eq:total lagrangian}).
Using the expressions for the conjugate momenta (\ref{eq:beta conjugate momentum})
and (\ref{eq:phi conjugate momentum}), the Hamiltonian is given by
\begin{eqnarray}
\mathcal{H} & = & -\frac{1}{2}\Pi_{i}^{\beta}\Pi_{\beta}^{i}-\frac{\xi}{2b_{0}^{4}}\Pi_{0}^{2}-\frac{b_{i}b^{i}}{2b_{0}^{2}}\left(\Pi_{0}^{\beta}\right)^{2}+\frac{b^{i}}{b_{0}}\Pi_{0}^{\beta}\Pi_{i}^{\beta}+\Pi_{\beta}^{i}\partial_{i}\beta_{0}-\frac{2b^{i}}{b_{0}}\Pi_{0}^{\beta}\partial_{i}\beta_{0}\nonumber \\
 &  & -\frac{b^{i}b^{j}}{b_{0}^{2}}\Pi_{0}^{\beta}\partial_{i}\beta_{j}-\frac{1}{4b_{0}^{2}}\Pi_{\phi}^{2}+\frac{1}{4}\beta_{ij}\beta^{ij}+\kappa b_{\mu}b_{\nu}\beta^{\mu}\beta^{\nu}-\frac{b^{i}}{b_{0}}\Pi_{\phi}\partial_{i}\phi+2\xi\kappa\phi^{2}.\label{eq:KSS hamiltonian}
\end{eqnarray}
As discussed before, there are five propagating degrees of freedom, gauge dependence,
and this Hamiltonian is unbounded from below. For this reason, one needs to
implement conditions on the states of the Fock space so that, when restricted to these states, all the mentioned problems can
be avoided. The extra degrees of freedom brought by the propagation of the gauge
modes are kept outside of the physical region by the imposition of the Gupta-Bleuler condition
(\ref{eq:gupta-bleuler condition}), which can be easily restated
using the expression for the conjugate momenta (\ref{eq:beta conjugate momentum})
as 
\begin{equation}
\left\langle \Pi_{\beta}^{0}+2b_{0}^{2}\sqrt{\kappa}\phi\right\rangle_{\text{Phys}} =0,\label{eq:ghost absence cond phase space}
\end{equation}
where the subscript ``Phys'' means that the expectation value is being calculated between physical states.

Considering the $\nu=0$ component of the Hamiltonian version of the equation of motion (\ref{eq:field eq beta}) for $\beta_{\mu}$
and using the condition above, one obtains
\begin{eqnarray}
\left\langle \Pi_{\phi}\right\rangle_{\text{Phys}}  & = & \left\langle \frac{1}{\sqrt{\kappa}}\partial_{i}\Pi_{\beta}^{i}+2b^{i}b_{0}\partial_{i}\phi-2\sqrt{\kappa}b_{\mu}b_{0}\beta^{\mu}\right\rangle_{\text{Phys}}.\label{eq:contraint phi mom}
\end{eqnarray}
These two constraints on the physical states suppress the dynamics
of two out of the five degrees of freedom, and $\left\langle \mathcal{H}\right\rangle_{\text{Phys}}$ can be conveniently written as 
\begin{eqnarray}
\left\langle \mathcal{H}\right\rangle_{\text{Phys}}  & = & \left\langle -\frac{1}{2}\left(\Pi^{i}_{\beta}+2\sqrt{\kappa}b_{0}b^{i}\phi\right)\left(\Pi_{i}^{\beta}+2\sqrt{\kappa}b_{0}b^{i}\phi\right)+\frac{1}{4}\beta_{ij}\beta^{ij}\right\rangle_{\text{Phys}} \nonumber \\
 &  & -\left\langle \left(\beta_{0}+\frac{1}{2\kappa b_{0}^{2}}\partial_{i}\Pi^{i}_{\beta}+\frac{2}{\sqrt{\kappa}b_{0}}b^{i}\partial_{i}\phi-\frac{1}{b_{0}}b_{\mu}\beta^{\mu}\right)\left(\frac{1}{2}\partial_{i}\Pi^{i}_{\beta}+\sqrt{\kappa}b^{0}b^{i}\partial_{i}\phi\right)\right\rangle_{\text{Phys}}.
\end{eqnarray}
First, one  notices the gauge independence of this expression as should be expected. To ensure the positiveness of this Hamiltonian is enough to require that
\begin{equation}
\left\langle \frac{1}{2}\partial_{i}\Pi^{i}_{\beta}+\sqrt{\kappa}b^{0}b^{i}\partial_{i}\phi\right\rangle_{\text{Phys}} =0.\label{eq:cond absence tackyons phase space}
\end{equation}
Interestingly, this expression could be obtained by introducing the Stueckelberg field, through substitution (\ref{eq:stueckelberg field}), into the classical stability condition $b\cdot\beta=0$ and considering it as a quantum expectation value between physical states. In this sense, the classical stability condition obtained in Ref. \cite{stability bumblebee} is recovered. 

From the expression (\ref{eq:beta conjugate momentum}) for the $\Pi_{\beta}^{\mu}$ field and using the constraint (\ref{eq:ghost absence cond phase space}), 
equation (\ref{eq:cond absence tackyons phase space}) can be written as $\left\langle\partial_{0}\partial_{i}\beta^{i}\right\rangle_{\text{Phys}} =\left\langle \partial_{i}\partial^{i}\beta^{0}\right\rangle_{\text{Phys}}$. Using the field expansion (\ref{eq:expansion complete field beta}) and after some algebraic manipulation,
this amounts to the condition $\left\langle d\left(\vec{q}\right)\right\rangle_{\text{Phys}} =0$,
which is ensured by the condition (\ref{eq:absence of tackyon condition}). To show the dynamic consistency of this condition, consider
the $\nu=i$ component of the Hamiltonian
version of the equation of motion (\ref{eq:field eq beta}) for $\beta_{\mu}$:
\begin{equation}
-\partial_{0}\Pi^{i}_{\beta}+\partial_{j}\beta^{ji}-\frac{b^{i}b^{j}}{b_{0}^{2}}\partial_{j}\Pi^{0}_{\beta}-2\kappa b_{\mu}b^{i}\beta^{\mu}=0.
\end{equation}
Multiplying this equation by $\partial_i$, taking the expectation value between physical states, and using the Gupta-Bleuler condition (\ref{eq:ghost absence cond phase space}), one obtains
\begin{equation}
\left\langle\partial_{i}\dot{\Pi}^{i}_{\beta}+\frac{b^{i}b^{j}}{b_{0}^{2}}\partial_{i}\partial_{j}\left(-2b_{0}^{2}\sqrt{\kappa}\phi\right)+2\kappa b_{\mu}b^{i}\partial_{i}\beta^{\mu}\right\rangle_{\text{Phys}}=0. \label{lema cons cond tac}
\end{equation}

The consistency of the condition (\ref{eq:cond absence tackyons phase space}) with the field dynamics is verified if the time derivative of the combination on the left-hand side of Eq. (\ref{eq:cond absence tackyons phase space}), $\left\langle \frac{1}{2}\partial_{i}\dot{\Pi}^{i}_{\beta}+\sqrt{\kappa}b^{0}b^{i}\partial_{i}\dot{\phi}\right\rangle_{\text{Phys}}$, vanishes. From Eq. (\ref{eq:phi conjugate momentum}), $\dot{\phi}=-\frac{1}{2b_{0}^{2}}\Pi_{\phi}-\frac{b_{i}}{b_{0}}\partial^{i}\phi$. Using this relation and rewriting $\Pi_{\phi}$ and $\partial_{i}\dot{\Pi}^{i}_{\beta}$ in terms of the expressions obtained from Eqs. (\ref{eq:contraint phi mom}) and (\ref{lema cons cond tac}), one finally gets
\begin{equation}
\left\langle \frac{1}{2b_{0}\sqrt{\kappa}}\partial_{i}\dot{\Pi}^{i}_{\beta}+b^{i}\partial_{i}\dot{\phi}\right\rangle_{\text{Phys}}=\left\langle -b^{j}\partial_{j}\left(\frac{1}{2b_{0}^{2}\sqrt{\kappa}}\partial_{i}\Pi^{i}_{\beta}+\frac{b^{i}}{b_{0}^{2}}\partial_{i}\phi\right)\right\rangle_{\text{Phys}},  
\end{equation}
which vanishes due to condition (\ref{eq:cond absence tackyons phase space}). Therefore, condition (\ref{eq:absence of tackyon condition})
is a stable one, and it ensures the absence of tachyons in the physical spectrum of the free theory.

With the analysis of this section, one concludes that the components to develop a systematic quantum analysis of bumblebee electrodynamics described by Lagrangian (\ref{eq:Lagrangian}) can be consistently constructed. Moreover, in spite of the fact that the present approach to the canonical quantization needs the introduction of unphysical ghost modes, a physical Fock space free from pathologies can be defined.

\section{Summary\label{sec:Summary}}

In this work, the problems of the quantization and
stability of a particular vector theory with a potential term that
triggers the spontaneous symmetry breaking of Lorentz symmetry were addressed. In
Sec. \ref{sec:Spontaneous-Lorentz-symetry}, the main classical
properties of the model described by the Lagrangian (\ref{eq:Lagrangian}) were reviewed.
Performing a Hamiltonian analysis, it was verified that the model exhibits
two second-class constraints and only three out of the four degrees of freedom available
in the vector field can be dynamical. Two of them correspond to the
massless NG modes and form a massless spin-1 particle
that, with the choice of the kinetic term, can be potentially identified
as the photon. The other propagating mode corresponds to a field excitation
that does not remain on the bottom of the potential, and for this reason
is characterized as a massive particle. In fact, the mass of this
particle was shown to be negative, leading to an instability in  the
model. However, at the end of that section it was shown that the instability
can be avoided if suitable initial conditions are chosen, and the reduced
phase space of this model is equivalent to that of the Maxwell electrodynamics
in a nonlinear gauge.

The construction of a framework
for discussing the quantum picture of the previous scenario was pursued in the subsequent sections. To
avoid some of the complications of the Dirac method of quantization of constrained systems, the known Stueckelberg trick was 
used. This consisted of promoting an enlargement
of the field content of the model with the simultaneous introduction of a local symmetry.
By doing this, the second-class constraints were converted to first-class ones, and the widely known and successful methods for the quantization of gauge theories could be applied.
In this vein, a convenient gauge-fixing term was added to the KS-Stueckelberg
Lagrangian, and physical equivalence with the KS model was claimed
based on the fact that the two Lagrangians differed by a BRST-invariant
term.

To discuss the perturbation theory analysis,  the free
Lagrangian (\ref{eq:total lagrangian}) was considered, and the Fourier decomposition of the free-field solutions was performed in Sec. \ref{sec:Fourier-modes-expansion}.
The dispersion relations of the propagating modes were obtained, 
and the appearance of two new unphysical degrees of freedom brought
by the gauge-fixing term was observed. The other three degrees of freedom were the expected
massless spin-$1$ and the tachyonic mode. In the sequence, the concept of pure-mode solutions was
introduced in Eq. (\ref{eq:pure mode solutions}), which
helped us to understand the structure of the Fourier decomposition.
These are particular solutions for the equations of motion, where the
$0$ component of the four-momentum that appears in the matrix operator
between brackets in Eq. (\ref{eq:pure mode solutions}) is one of
the solutions, $p_{0}^{(\lambda)}$, for the dispersion relations 
(\ref{eq:disp rel beta gauge dep})\textendash(\ref{eq:disp rel tachyon}).
In some cases, like in the quantization of the Maxwell electrodynamics
modified by the introduction of the gauge-fixing and finite-mass terms
and in the photon sector of the SME, these pure-mode solutions form a
basis of vectors for each value of the three-momentum $\vec{p}$, and
they provide a natural set of polarization vectors appearing
in the Fourier decomposition. However, in the present discussion,
the pure-mode solution associated with the $\lambda=0$ dispersion relation changes its spacelike, timelike or lightlike behavior for different choices
of the $3$-momentum. It turns out that the four pure-mode solutions do
not provide a basis of vectors for every choice of the three-momentum.
This makes the Fourier decomposition of the vector field
much more involved, as can be seen from the final result (\ref{eq:expansion complete field beta}).

With the Fourier decomposition of the Stueckelberg and vector fields, the construction of the physical
Fock space of the model was
discussed in Sec. \ref{sec:Stability}. The Fourier expansion
of the fields was inverted and by using the ETCR (\ref{eq:phi com rel})\textendash(\ref{eq:beta dot beta dot com rel}),
the algebra of the Fourier-mode operators was obtained. In the transverse
sector, these operators could be directly interpreted as creation and
annihilation operators of massless spin-$1$ particles. Due to the
sign in the right-hand side of Eq. (\ref{eq:c algebra}), one could also verify
that the full Fock space is plagued by negative-norm states. These
ghost states are commonly introduced in the quantization of gauge theories,
and they are excluded from the physical Fock space by demanding that
the extra gauge modes, brought by the gauge-fixing terms, be kept
out of the physical Fock space. This was attained, in the present case, by imposing that
a physical state should satisfy the condition (\ref{eq:gupta-bleuler condition}).
The information about the tachyon mode is contained in the $s$ and
$d$ operators, but a condition for the absence of tachyons in the physical
Fock space cannot be obtained directly from them, since they cannot
be interpreted as creation or annihilation operators without contradicting
their algebraic relations. The proposed solution for this problem was
to redefine them in terms of the operators $\tau$ and $\tau^{\dagger}$
in such a way that these two new operators carried the same information
as the previous ones and satisfied a usual creation and annihilation
operator algebra. So, it was finally proposed that the condition
(\ref{eq:absence of tackyon condition}) is the one that guarantees
the absence of tachyons in the physical spectrum and, therefore, the
stability of the free model. The same condition was regained by demanding the positiveness of the physical
free Hamiltonian. Then, the stability condition  was restated  in terms of the fields which, in turn, was verified to be the quantum analogous to the classical stability condition
discussed in Sec. \ref{sec:Spontaneous-Lorentz-symetry} if the redefinition of the fields by the Stueckelberg method is taken into account.

The main attainment of this work was to construct the building blocks to develop a systematic quantum analysis of the KS model. This was achieved by showing that one can define a region of the full Hilbert space where
the free model is stable and is, indeed, equivalent to the Maxwell electrodynamics
in the temporal gauge. The present framework paves the road  for further discussions of the quantum properties of the KS model that are presumably of great importance but lie beyond the scope of the present work, like the stability of the physical Fock space under radiative corrections, the coupling to the matter sector, and microcausality-related issues.

\section*{Acknowledgments}

The author gratefully acknowledges V. Alan Kosteleck\'{y} for the useful discussions and Robert Bluhm and Robertus Potting for carefully reading the manuscript of the present work and for sending useful comments. This work has been supported by CAPES
(Coordena\c{c}\~{a}o de Aperfei\c{c}oamento de Pessoal de N\'{\i}vel Superior-Brazil).

\appendix

\section{BRST INVARIANCE \label{appendix a}}

The BRST invariance of the full Lagrangians $\mathcal{L}_{KSS}$ and $\mathcal{L}_{KSS}+\mathcal{L}_{gf}$ will be shown in this appendix.

We introduce the scalar anticommutating fields $\omega(x)$ and $\omega^{\ast}(x)$. These are called ghost fields, since they satisfy a wrong spin statistics relation. From the invariance of the free Lagrangian $\mathcal{L}_{KSS}$ under the gauge transformations (\ref{eq:gauge transf beta}) and (\ref{eq:gauge transf phi}), it can be seen that it is also invariant under the particular infinitesimal gauge transformations
\begin{eqnarray}
{\bf s}\beta_{\mu}(x)&=&\partial_{\mu}\omega(x),\\
{\bf s}\phi(x)&=&\sqrt{\kappa}\omega(x),\\
{\bf s}\omega(x)&=&0,
\end{eqnarray}
where ${\bf s}$ is the operator that performs the infinitesimal BRST transformations. Since $\omega(x)$ and $\omega^{\ast}(x)$ are anticommutating fields, ${\bf s}$ is nilpotent, ${\bf s}^2=0$.

Instead of the gauge-fixing Lagrangian (\ref{eq:gauge fixing Lagrangian}), consider the more general one
\begin{equation}
\mathcal{L}_{gf}={\bf s}\left[\omega^{\ast}\left(\mathcal{G}(\beta_{\mu},\phi)+\frac{\xi}{2}h\right)\right], \label{general gf}
\end{equation}
where $\mathcal{G}$ is some general functional of the fields $\beta_{\mu}$ and $\phi$, and $h$ is an auxiliary field called the Nakanishi-Lautrup field. Whatever the particular form of the functional 
$\mathcal{G}$, this Lagrangian is BRST invariant if the following transformations for the $\omega^{\ast}$ and $h$ fields are assumed:
\begin{eqnarray}
{\bf s}\omega^{\ast}(x)&=&h,\\
{\bf s}h&=&0.
\end{eqnarray}
Notice that {\bf s} obeys the Leibniz product rule, and it anticommutes with the ghost fields. Deriving the algebraic equation of motion for the auxiliary $h$ field and using it to eliminate this field from the Lagrangian (\ref{general gf}), one obtains
\begin{equation}
\mathcal{L}_{gf}=-\omega^{\ast}\left({\bf s}\mathcal{G}\right)-\frac{1}{2\xi}\mathcal{G}^2.
\end{equation}
Choosing $\mathcal{G}=\left(b^{\nu}b^{\mu}\partial_{\nu}\beta_{\mu}-2\xi\sqrt{\kappa}\phi\right)^{2}$, one can write
\begin{equation}
\mathcal{L}_{gf}=-\omega^{\ast}\left(b^{\nu}b^{\mu}\partial_{\nu}\partial_{\mu}-2\xi\sqrt{\kappa}\right)\omega-\frac{1}{2\xi}\left(b^{\nu}b^{\mu}\partial_{\nu}\beta_{\mu}-2\xi\sqrt{\kappa}\phi\right)^{2}.
\end{equation}

Therefore, with this choice for the functional $\mathcal{G}$, the ghost fields decouple and can be discarded from the Lagrangian without affecting physical results. In this way, the BRST invariance of the total Lagrangian $\mathcal{L}_{KSS}+\mathcal{L}_{gf}$ is established.

\section{PARTICULAR SOLUTION OF Eq. (\ref{eq:eq motion mom long mode})  \label{appendix b}}

In this appendix, the particular
solution for the inhomogeneous differential equation (\ref{eq:eq motion mom long mode}) is derived .
The problem is essentially that the naive convolution of the Green function with the inhomogeneous piece coming from (\ref{eq:fourier decomp beta along b}) will lead to the
product $\frac{1}{p\cdot b}\delta\left(p\cdot b\right)$, which is
ill defined. However, the issue only abides in the convolution of
the Green function with the $d$ term in Eq. (\ref{eq:fourier decomp beta along b})
that is where $\delta\left(p\cdot b\right)$ plays a role. There is
no prevention in forming the convolution of the Green function with
the $c$ and $c^{\dagger}$ terms.

The particular solution can be derived by splitting
it into two pieces: 
\begin{equation}
S^{P}=S_{1}^{P}+S_{2}^{P},\label{eq:part sol div beta 2 parts}
\end{equation}
where $S_{1}^{P}$ is the particular solution for the equation
\begin{equation}
\left(b\cdot\partial\right)S_{1}^{P}\left(x\right)=\Box C\left(x\right) \label{eq:sol part 1 div beta}
\end{equation}
and $C\left(x\right)$ is the function defined by the first integral
on the right-hand side of Eq. (\ref{eq:div beta particular}). That is, 
\begin{equation}
C\left(x\right)=-\int\frac{d^{3}p}{4\kappa\left|b_{0}\right|\left(2\pi\right)^{3}}
\left(c\left(\vec{p}\right)e^{-ip^{(0)}\cdot x}+c^{\dagger}\left(\vec{p}\right)e^{ip^{(0)}\cdot x}\right).\label{eq:def function C}
\end{equation}
The other piece of the particular solution in Eq. (\ref{eq:part sol div beta 2 parts})
is a particular solution for the equation
\begin{equation}
\left(b\cdot\partial\right)S_{2}^{P}\left(x\right)=\left(\Box-2\kappa b^{2}\right)D\left(x\right),\label{eq:sol part 2 div beta}
\end{equation}
where $D\left(x\right)$ is the function defined by the second integral
on the right-hand side of Eq. (\ref{eq:div beta particular}). That is,
\begin{equation}
D\left(x\right)=\int\frac{d^{3}p}{\left(2\pi\right)^{3}\left|b_{0}\right|}d\left(\vec{p}\right)e^{-ip^{\left(3\right)}\cdot x}.\label{eq:def function D}
\end{equation}
It can be easily verified that the two functions defined as particular
solutions in Eqs. (\ref{eq:sol part 1 div beta}) and (\ref{eq:sol part 2 div beta}),
when added, form a particular solution to Eq. (\ref{eq:eq motion mom long mode}),
since
\begin{equation}
\left(\left(b\cdot\partial\right)^{2}-2\kappa\xi\right)C\left(x\right)=0
\end{equation}
and
\begin{equation}
\left(b\cdot\partial\right)D\left(x\right)=0.\label{eq:dif eq D}
\end{equation}

To obtain $S_{1}^{P}$, one considers the Green function for the operator
$b\cdot\partial$, which must satisfy
\begin{equation}
\left(b\cdot\partial\right)G\left(x\right)=\delta\left(x\right).
\end{equation}
Since it is only needed for a particular solution, the following
Green function that satisfies this equation is chosen :
\begin{equation}
G^{\left(3\right)}\left(x\right)=\frac{i}{\left(2\pi\right)^{4}}\int d^{4}p\frac{e^{-ip\cdot x}}{b\cdot p+i\epsilon}.\label{eq:green function 3}
\end{equation}
Now, the particular solution for Eq. (\ref{eq:sol part 1 div beta})
can be written as
\begin{equation}
S_{1}^{P}\left(x\right)=\Box\left(G^{\left(3\right)}\ast C\right)\left(x\right),\label{eq:sol part 1 conv}
\end{equation}
where the symbol ``$\ast$'' means convolution, which for two arbitrary
functions $f_{1}\left(x\right)$ and $f_{2}\left(x\right)$ is defined
by
\begin{equation}
f_{1}\ast f_{2}\left(x\right)=\int d^{4}yf_{1}\left(x-y\right)f_{2}\left(y\right).
\end{equation}
The property of the convolution operation:
\begin{equation}
\mathcal{O}\left(f_{1}\ast f_{2}\right)=\left(\mathcal{O}f_{1}\ast f_{2}\right)=\left(f_{1}\ast\mathcal{O}f_{2}\right),
\end{equation}
where $\mathcal{O}$ is a derivative operator, was also used in Eq. (\ref{eq:sol part 1 conv}).

As was observed before, one cannot apply the same technique for
the obtainment of $S_{2}^{P}$, since the convolution $G^{\left(3\right)}\ast D$
is ill defined. To proceed with this calculation, one considers the
continuous deformation, $D\left(x;\tau\right)$, of the function $D\left(x\right)$
that analogously to Eq. (\ref{eq:dif eq D}) satisfies the differential
equation
\begin{equation}
\left(b\cdot\partial+\tau\right)D\left(x;\tau\right)=0,
\end{equation}
with the subsidiary condition that $D\left(x;\tau\right)\longrightarrow D\left(x\right)$
when $\tau\longrightarrow0$. This gives
\begin{equation}
D\left(x;\tau\right)=\int\frac{d^{3}p}{\left(2\pi\right)^{3}|b_{0}|}d\left(\vec{p}\right)e^{-i\left(\frac{\vec{b}\cdot\vec{p}}{b_{0}}-i\frac{\tau}{b_{0}}\right)x^{0}+i\vec{p}\cdot\vec{x}}\label{eq:deformed D}
\end{equation}

Similarly, $S\left(x;\tau\right)$ is defined in such a way that it satisfies the following
differential equations:
\begin{eqnarray}
\left(b\cdot\partial\right)S_{2}^{P}\left(x;\tau\right) & = & \left(\Box-2\kappa b^{2}\right)D\left(x;\tau\right),\\
\left(b\cdot\partial+\tau\right)S_{2}^{P}\left(x;\tau\right) & = & \left(\Box-2\kappa b^{2}\right)D\left(x\right).
\end{eqnarray}
Subtracting the first of these equations from the second, dividing
both sides by $\tau$, and taking the limit $\tau\longrightarrow0$,
one gets
\begin{equation}
S_{2}^{P}\left(x\right)=\left(\Box-2\kappa b^{2}\right)\frac{dD\left(x;\tau\right)}{d\tau}\bigg|_{\tau=0}.
\end{equation}
Gathering $S_{1}^{P}$ and $S_{2}^{P}$ along with the
homogenous solution (\ref{eq:div beta homogeneous}), one has the desired general solution for Eq. (\ref{eq:sol part 1 div beta}).
Using Eqs. (\ref{eq:green function 3}) and (\ref{eq:deformed D}), Eq. (\ref{eq:div beta particular}) is obtained.


\begin{thebibliography}{9}

\bibitem{string1} V. A. Kosteleck\'{y} and S. Samuel, Phys. Rev. D \textbf{39}, 683 (1989); Phys. Rev.
Lett. \textbf{63}, 224 (1989).

\bibitem{string2} V. A. Kosteleck\'{y} and S. Samuel, Phys. Rev. Lett. \textbf{66}, 1811 (1991);
V. A. Kosteleck\'{y} and R. Potting, Nucl. Phys. \textbf{B359}, 545 (1991);
Phys. Lett. B \textbf{381}, 89 (1996); Phys. Rev. D \textbf{63}, 046007 (2001);
V. A. Kosteleck\'{y}, M. J. Perry, and R. Potting, Phys. Rev. Lett. \textbf{84}, 4541 (2000).

\bibitem{lqg} J. Alfaro, H. A. Morales-T´ecotl, and L. F. Urrutia, Phys. Rev. Lett. \textbf{84}, 2318 (2000);
Phys. Rev. D \textbf{65}, 103509 (2002).

\bibitem{non comm} See, for example, I. Mocioiu, M. Pospelov, and R. Roiban, Phys. Lett. B \textbf{489}, 390 (2000);
S. M. Carroll, J. A. Harvey, V. A. Kosteleck\'{y}, C. D. Lane, and T. Okamoto, Phys. Rev. Lett. \textbf{87}, 141601 (2001);
Z. Guralnik, R. Jackiw, S. Y. Pi, and A. P. Polychronakos, Phys. Lett. B \textbf{517}, 450 (2001);
C. E. Carlson, C. D. Carone, and R. F. Lebed, Phys. Lett. B \textbf{518}, 201 (2001);
A. Anisimov, T. Banks, M. Dine, and M. Graesser, Phys. Rev. D \textbf{65}, 085032 (2002);
A. Das, J. Gamboa, J. Lopez-Sarrion, and F. A. Schaposnik, Phys. Rev. D \textbf{72}, 107702 (2005).

\bibitem{non trivial top} F.R. Klinkhamer, Nucl. Phys. \textbf{B578}, 277 (2000).

\bibitem{coll-kost} D. Colladay and V. A. Kosteleck\'{y}, Phys. Rev. D \textbf{58}, 116002 (1998).

\bibitem{alan-gravity} V. A. Kosteleck\'{y}, Phys. Rev. D \textbf{69}, 105009 (2004).

\bibitem{SSB lorentz} V. A. Kosteleck\'{y} and S. Samuel, Phys. Rev. D \textbf{40}, 1886 (1989)

\bibitem{colladay-kost} D. Colladay and V. A. Kosteleck\'{y}, Phys. Rev. D \textbf{55}, 6760 (1997).

\bibitem{gravity} Q. G. Bailey and V. A. Kosteleck\'{y}, Phys. Rev. D \textbf{74}, 045001 (2006).

\bibitem{bluhm} R. Bluhm and V. A. Kosteleck\'{y}, Phys. Rev. D \textbf{71}, 065008 (2005).


\bibitem{schreck manoel} R. Casana, M. M. Ferreira, A. R. Gomes, and P. R. D. Pinheiro, Phys. Rev. D \textbf{80}, 125040 (2009);
R. Casana, M. M. Ferreira, A. R. Gomes, and F. E. P. dos Santos, Phys. Rev. D \textbf{82}, 125006 (2010);
F. R. Klinkhamer and M. Schreck, Nucl. Phys. \textbf{B848}, 90 (2011); Nucl. Phys. \textbf{B856}, 666 (2012); 
M. Schreck, Phys. Rev. D \textbf{86}, 065038 (2012); Phys. Rev. D \textbf{89}, 085013 (2014); Phys. Rev. D \textbf{89}, 105019 (2014).

\bibitem{alan-lehnert}  V. A. Kosteleck\'{y} and R. Lehnert, Phys. Rev. D \textbf{63}, 065008 (2001).

\bibitem{stability bumblebee} R. Bluhm, N. L. Gagne, R. Potting, and A. Vrublevskis, Phys. Rev. D \textbf{77}, 125007 (2008).
 
\bibitem{bumblebee} B. Altschul and V. A. Kosteleck\'{y}, Phys. Lett. B \textbf{628}, 106 (2005);
V. A. Kosteleck\'{y} and J. D. Tasson, Phys. Rev. D \textbf{83}, 016013 (2011).

\bibitem{nambu} Y. Nambu, Prog. Theor. Phys. Suppl. Extra Number, \textbf{190} (1968).

\bibitem{fung} R. Bluhm, Shu-Hong Fung, and V. A. Kosteleck\'{y}, Phys. Rev. D \textbf{77}, 065020 (2008).

\bibitem{massive vector} J. H. Lowenstein and B. Schroer, Phys. Rev. D \textbf{6}, 1553 (1972); H. van Hees, arXiv:0305076.

\bibitem{altaba} H. Ruegg and M. Ruiz-Altaba,  Int. J. Mod. Phys. A \textbf{19}, 3265 (2004).

\bibitem{lee} B. Lee, ``Gauge Theories,'' in {\it Methods in Field Theory}, edited by R. Balain and J. Zinn-Justin (Les Houches, France 1976).

\bibitem{gabadadze} G. Gabadadze and L. Grisa, Phys. Lett. B \textbf{617}, 124 (2005).

\bibitem{dvali} G. Dvali, M. Papucci, and M. D. Schwartz, Phys. Rev. Lett. \textbf{94}, 191602 (2005).

\bibitem{altschul} B. Altschul, Phys. Rev. D \textbf{73}, 036005 (2006).

\bibitem{potting gupta} D. Colladay, P. McDonald, and R. Potting, Phys. Rev. D \textbf{89}, 085014 (2014).



\end{thebibliography}
\end{document}